\documentclass[journal,twoside,web, letterpaper]{ieeecolor}
\usepackage{tac}
\let\labelindent\relax
\def\tacversion{1}

\usepackage{cite}[sort,compress]
\usepackage[pdftex]{graphicx}
\usepackage{array}
\usepackage{url}
\usepackage{arydshln}
\usepackage{float}
\usepackage{scrextend}
\usepackage{blindtext}
\usepackage{amsmath,amsfonts,amssymb}
\usepackage{tikz}
\usepackage{xspace}
\usepackage{xcolor}
\usepackage{mathrsfs,mathtools}
\usepackage{mathalfa}
\usepackage{euscript}
\usepackage{csquotes}
\usepackage{datetime}
\usepackage{arydshln}
\usepackage[inline]{enumitem}
\usepackage[normalem]{ulem}
\usepackage{algorithm}

\usepackage{algpseudocode}
\algrenewcommand\algorithmicrequire{\textbf{Requires}}
\algrenewcommand\algorithmicensure{\textbf{Outputs}}
\usepackage{subcaption}

\definecolor{mblue}{rgb}{0,0.4470,0.7410}
\definecolor{morange}{rgb}{0.8500,0.3250,0.0980}
\definecolor{myellow}{rgb}{0.9290,0.6940,0.1250}
\definecolor{mpurple}{rgb}{0.4940,0.1840,0.5560}
\definecolor{mgreen}{rgb}{0.4660,0.6740,0.1880}
\definecolor{mcyan}{rgb}{0.3010,0.7450,0.9330}
\definecolor{mred}{rgb}{0.6350,0.0780,0.1840}
\definecolor{mgreenblue}{rgb}{0.0,1.0,0.5}
\definecolor{parulablue}{rgb}{0.2431,0.1490,0.6588}
\definecolor{parulalblue}{RGB}{39,151,235}
\definecolor{parulagreen}{RGB}{129,204,89}
\definecolor{parulayellow}{RGB}{249,251,21}
\definecolor{cblue}{rgb}{0,0.9,1}
\definecolor{corange}{rgb}{1,0.7,0}
\definecolor{mgray}{rgb}{0.8,0.8,0.8}

\ifx\tacversion\undefined%
    \newenvironment{definition}{\begin{defn}}{\end{defn}}
    \newenvironment{problem}{\begin{prob}}{\end{prob}}
    \newenvironment{proposition}{\begin{prop}}{\end{prop}}
    \newenvironment{theorem}{\begin{thm}}{\end{thm}}
    \newenvironment{lemma}{\begin{lem}}{\end{lem}}
    \newenvironment{corollary}{\begin{cor}}{\end{cor}}
    \newenvironment{remark}{\begin{rem}}{\end{rem}}
    \newenvironment{condition}{\begin{cond}}{\end{cond}}
    \newenvironment{proof}{\begin{pf}}{\hfill$\blacksquare$\end{pf}}
    
    \newcommand{\makebiography}[3]{
    \renewcommand{\baselinestretch}{0.83}
    \selectfont
    \par\noindent 
    \parbox[t]{\linewidth}{
    \noindent\parpic{\includegraphics[width=.85in,keepaspectratio]{#1}}
    \noindent {\scriptsize  {\bf #2}\
    #3}}}
    
\else%
    \usepackage{amsthm}
    \theoremstyle{definition}
    \newtheorem{definition}{Definition}
    \newtheorem{exmp}{Example}
    \newtheorem{problem}{Problem}
    \theoremstyle{plain}
    \newtheorem{theorem}{Theorem}

    \newtheorem{lemma}{Lemma}
    
    \newtheorem{proposition}{Proposition}

    \theoremstyle{remark}

    \newcommand{\makebiography}[3]{
        \begin{IEEEbiography}%
        [{\includegraphics[width=1in,height=1.25in,clip,keepaspectratio]{#1}}]%
        {#2}
            #3
        \end{IEEEbiography}%
    }
\fi%

\newenvironment{example}{\begin{exmp}}{\hfill$\blacktriangleleft$\end{exmp}}

\newcommand{\tss}[1]{\textsuperscript{#1}}
\newcommand{\lpvcore}{\textsc{LPVcore}\xspace}

\newcommand{\comment}[1]{}
\newcounter{ass}

\newcommand{\mc}[1]{\mathcal{#1}}
\newcommand{\mf}[1]{\mathfrak{#1}}
\newcommand{\mr}[1]{\mathrm{#1}}
\newcommand{\mb}[1]{\mathbb{#1}}
\newcommand{\ms}[1]{\mathscr{#1}}

\newcommand{\mt}[1]{\mathtt{#1}}
\newcommand{\mbf}[1]{\mathbf{#1}}
\newcommand{\meu}[1]{\EuScript{#1}}
\DeclareMathAlphabet{\mathpzc}{OT1}{pzc}{m}{it}

\newcommand{\unaryminus}{\scalebox{0.65}[1]{\ensuremath{\,-}}}

\newcommand{\rank}{\mr{rank}}
\newcommand{\col}{\mr{col}}

\newcommand{\kron}{\otimes} %
\newcommand{\bkron}{\circledcirc}

\newcommand{\dnx}{{n_\mr{x}}}
\newcommand{\dny}{{n_\mr{y}}}
\newcommand{\dnu}{{n_\mr{u}}}
\newcommand{\dnp}{{n_\mr{p}}}
\newcommand{\dnw}{{n_\mr{w}}}
\newcommand{\dna}{{n_\mr{a}}}
\newcommand{\dnb}{{n_\mr{b}}}
\newcommand{\dnr}{{n_\mr{r}}}
\newcommand{\dnk}{{n_\mr{k}}}

\newcommand{\svdots}{\raisebox{0pt}{$\scalebox{.75}{\vdots}$}}
\newcommand{\sddots}{\raisebox{0pt}{$\scalebox{.75}{$\ddots$}$}}

\newcommand{\weight}[2][]{%
  \mf{W}_{\mathtt{#2}}%
  \if\relax\detokenize{#1}\relax
  \else
    ^{^{[#1]}}%
  \fi
}

\newcommand{\q}{\mr{q}}

\newcommand{\nBf}{\mbf{n}(\mf{B})}
\newcommand{\mBf}{\mbf{m}(\mf{B})}
\newcommand{\cBf}{\mbf{c}(\mf{B})}
\newcommand{\mBpf}{\mbf{m}(\mf{B}^\prime)}

\newcommand{\LBf}{\mbf{L}(\mf{B})}

\newcommand{\Nd}{{N_\mr{d}}}
\newcommand{\Bfint}[2]{\left.\mf{B}_{#1}\right|_{#2}}
\newcommand{\dataset}{\mc{D}_\Nd}

\newcommand{\Bss}{\mf{B}^{{\normalfont\textrm{\textsc{ss}}}}}

\newcommand{\springconstant}{s}

\newdateformat{monthyeardate}{\monthname[\THEMONTH], \THEYEAR}
\begin{document}
\title{A behavioral approach for LPV data-driven representations}
\author{%
Chris Verhoek,
Ivan Markovsky,
Sofie Haesaert, and
Roland T{\'o}th
\thanks{This work has been supported by The MathWorks Inc. and by the European Union within the framework of the National Laboratory for Autonomous Systems (RRF-2.3.1-21-2022-00002). Opinions, findings, conclusions or recommendations expressed in this paper are those of the authors and do not necessarily reflect the views of the MathWorks Inc. or the European Union.}
\thanks{C. Verhoek, R. T\'oth and S. Haesaert are with the Control Systems Group, Eindhoven University of Technology, The Netherlands. I. Markovsky is with the Catalan Institution for Research and Advanced Studies (ICREA), Barcelona, and the International Centre for Numerical Methods in Engineering (CIMNE), Barcelona, Spain.
R. T\'oth is also with the {HUN-REN} Institute for Computer Science and Control, Hungary. Email addresses: \textbraceleft \texttt{c.verhoek}, \texttt{r.toth}, \texttt{s.haesaert}\textbraceright \texttt{@tue.nl}, and \texttt{imarkovsky@cimne.upc.edu}. Corresponding author: Chris Verhoek.}
}

\maketitle
\begin{abstract}
	In this paper, we present a data-driven representation for \emph{linear parameter-varying} (LPV) systems, which can be used for direct data-driven analysis and control of such systems. Specifically, we use the behavioral approach to develop a data-driven representation of the finite-horizon behavior of LPV systems for which there exists a kernel representation with \emph{shifted-affine} scheduling dependence. Moreover, we provide a necessary and sufficient rank-based test on the available data that concludes whether the data \emph{fully} represents the finite-horizon {LPV} behavior. Using {the proposed data-driven} representation, we also solve the data-driven simulation problem {for LPV systems}. {Through {multiple} examples, we demonstrate that the results in this paper allow us {to formulate a novel set of} direct data-driven analysis and control {methods for} LPV systems, which are also applicable for LPV embeddings of nonlinear systems.}

\end{abstract}
\section{Introduction}\label{s:int}

Due to the ever-growing complexity of engineering systems, hindering traditional modeling methods, and the increasing availability of data
there is a growing 
interest in accomplishing analysis and control design directly on the basis of data. Particularly, direct data-driven analysis and control methods founded on the behavioral framework have gained a lot of attention. This is because these methods allow for %
rigorous stability and performance guarantees. A key result %
that enables using such approaches is
Willems' Fundamental Lemma~\cite{WillemsRapisardaMarkovskyMoor2005}, which allows for representing the behavior of a \emph{discrete-time} (DT) \emph{linear time-invariant} (LTI) system using a single sequence of measurement data, where the input is \emph{persistently exciting} (PE) of a certain order, i.e., the data is sufficiently `rich'. Based on this result, many data-driven analysis and control methods have been developed for DT LTI systems, see, e.g.,~\cite{markovsky2021behavioral} for an overview. {%
{A further generalization of the PE condition is the}
so-called \emph{generalized persistence of excitation} (GPE) condition~\cite{markovsky2022identifiability}, %
which provides an
(a posteriori) 
{test to decide} whether the \emph{data} is rich enough {to represent the underlying behavior of the system}. In the literature, `the Fundamental Lemma' is generally used to refer to the data-driven representation of the finite-horizon behavior, irrespective of whether it is {used in conjunction with} the PE condition or the GPE condition.}

Many extensions of the Fundamental Lemma have been proposed in the literature, such as  for \emph{continuous-time} (CT) LTI systems~\cite{rapisarda2023orthogonal}, 2D LTI systems~\cite{rapisarda2024input}, %
stochastic LTI systems~\cite{pan2022stochastic}, and variants for convex, conical, and affine behaviors~\cite{padoan2023data}. Beyond the class of LTI systems, %
several approaches {have been proposed to} %
extend the Fundamental Lemma towards specific classes of nonlinear systems, e.g.,~\cite{nortmann2023direct, alsalti2023data, fazzi2023data, rueda2020data}. These methods, however, generally impose restrictive assumptions on the system {class} (e.g., the system must be feedback linearizable, periodic, or described by a {finite} Volterra series). These assumptions allow for recasting the problem %
{in an}
equivalent 
{LTI form on which the standard} %
PE or GPE conditions {can be applied}.

{Alternatively, an} %
efficient way {of handling} nonlinear systems is via the framework of \emph{linear parameter-varying} (LPV) systems. {For} LPV systems,  %
the behavior is defined by a linear {dynamic} relationship that depends on a \emph{measurable} signal~$p$, called the \emph{scheduling signal}~{\cite{verhoek2024encyclo, shamma2012overview}}. {Such LPV representations} %
{are} often used {as} linear, surrogate models for the analysis and control of nonlinear systems, where the scheduling signal captures the nonlinearities and exogenous {effects~\cite{Toth2010, hoffmann2014survey}. This makes the LPV framework a bridge between linear and nonlinear analysis and control~\cite{shamma2012overview}. In fact, the LPV framework has been successfully applied in practice to solve complex nonlinear control problems~\cite{wassink2005lpv, nogueira2018lpv, kwon2011pointing}. We refer to~\cite{verhoek2024encyclo} for a broad, {up-to-date} overview of the current state-of-the-art in (model-based) LPV analysis and control.} Due to the {attractive properties and success of the LPV framework,}
developing direct data-driven analysis and control methods for LPV systems is an important stepping stone to achieve a generalization of the original {LTI} data-driven results %
to the nonlinear case. In this paper, we {aim to accomplish this in terms of} %
generalization of the {behavioral} data-driven representation for the class of LPV systems {and developing a} %
corresponding GPE condition. %

{To formulate our results, we} will %
{restrict the scope to}
LPV systems whose behavior can be characterized by an LPV kernel representation that has \emph{shifted-affine} scheduling dependence. We refer to this subclass %
 as LPV-SA systems. This is a highly useful subclass, as {these} systems admit a direct LPV \emph{state-space}~(SS) representation with static scheduling dependence, which is convenient for usage in LPV analysis and control methods. Moreover, through the 2\tss{nd} Fundamental Theorem of Calculus, a large class of nonlinear systems\footnote{{Specifically, the class of nonlinear {DT} systems with a continuously differentiable input-output map. That is, for $y(k) = f(y(k-1), \dots, y(k-n),u(k),\dots, u(k-n))$, with $f\in\mc{C}_1$.}} can be modeled {in an} LPV-SA {form}~\cite{koelewijn2023thesis, verhoek2024kernelnote}. 
 
In~\cite{Verhoek2021_CDC}, a rather general extension of the Fundamental Lemma was given for the class of LPV systems with meromorphic, dynamic scheduling dependence. {{However, beyond the theoretical interest,} this formulation %
results in a data-driven representation that {uses} %
composition weights {for the data that are arbitrary meromorphic functions of the scheduling signal with an arbitrary number of lags}. As there are no systematic
methods available that can compute these meromorphic weights, the data-driven LPV representation in~\cite{Verhoek2021_CDC} is {difficult to apply} in practice.} It was also shown as a remark in~\cite{Verhoek2021_CDC}, that this general form of the LPV Fundamental Lemma can be reduced to a simpler, practically useful form for LPV-SA systems, based on which a whole series of contributions on data-driven LPV methods has been developed. {Particularly, direct data-driven predictive control schemes~\cite{VerhoekAbbasTothHaesaert2021, verhoek2023linear}, direct data-driven LPV state-feedback control synthesis~\cite{Verhoek2022_DDLPVstatefb}, direct data-driven dissipativity analysis~\cite{Verhoek2023_dissipativity}, or efficient identification schemes~\cite{markovskyVerhoek2024}.} However, the LPV Fundamental Lemma for this rather useful subclass of LPV-SA systems has never been directly formulated. Furthermore, {\emph{computable}} conditions to check whether the data is `rich' enough, i.e., {the derivations of} GPE conditions, {have} never {been} sorted out. Moreover, the LPV data-driven simulation problem has not been formally solved yet. In this paper, we fill these gaps in the current literature {by providing the following contributions:} %
\begin{enumerate}[label={C\arabic*:}, align=left, ref={C\arabic*}, leftmargin=*]
    \item We formulate a finite-horizon data-driven LPV representation for LPV-SA systems that is {directly and easily} computable from a given data set;\label{C:rep}
    \item We provide a necessary and sufficient condition that is verifiable {on} the given data set to conclude whether the data can characterize the full finite-horizon behavior of the LPV-SA system;\label{C:cond}
    \item We provide a formal solution to the  data-driven simulation problem;\label{C:sim}
    \item {We demonstrate the capabilities of our results in terms of {solving} simulation and control problems on an LPV system and a nonlinear system.} \label{C:exmp}
\end{enumerate}
The remainder of the paper is structured as follows. We formalize the {considered} problem {setting} in Section~\ref{s:prob}. Section~\ref{s:sap} discusses the properties of behaviors of LPV-SA systems, such as their complexity and dimension. The data-driven representation and associated conditions on the data, constituting to Contributions~\ref{C:rep} and~\ref{C:cond}, are given in Section~\ref{s:main}. The formalization of the simulation problem and its solution (Contribution~\ref{C:sim}) are given in Section~\ref{s:sim}, while Section~\ref{s:examples} presents two examples that demonstrate the applicability of our methods in analysis and control {of} LPV and nonlinear systems. {Finally,} conclusions and possible future research directions are given in Section~\ref{s:con}.

\subsection*{Notation:}
$\mb{R}$ is the set of real numbers, while the set of integers is given by $\mb{Z}$. Consider the subspaces $\mb{A},\mb{B}$. The projection of $\mb{D}\subseteq\mb{A}\times\mb{B}$ onto the elements of $\mb{A}$ is denoted by $\pi_\mr{a}\mb{D} = \{a \in\mb{A}\mid(a,b)\in\mb{D}\}$, while $\mb{B}^\mb{A}$ indicates the collection of all maps from $\mb{A}$ to $\mb{B}$. We denote the dimension of a subspace by $\mr{dim}(\mb{A})$. The $p$-norm of a vector $x\in\mathbb{R}^{n_\mathrm{x}}$ is denoted by $\lVert x\rVert_p$. For the two matrices $A\in\mb{R}^{n\times m}$ and $B\in\mb{R}^{p\times q}$, the Kronecker product is given as $A\kron B\in\mb{R}^{pm \times qn}$, while $\mathrm{blkdiag}$ is the block-diagonal %
{composition}
for matrices, i.e., $\mathrm{blkdiag}(A,B) = \begin{bsmallmatrix} A & 0\\ 0 & B \end{bsmallmatrix}\in\mb{R}^{n+p\times m+q}$. The identity matrix of size $n\times n$ is denoted as $I_n$. Furthermore, $\col(x_1, \dots ,x_n)$ denotes the column vector $\begin{bsmallmatrix}x_1^\top & \cdots & x_n^\top \end{bsmallmatrix}^\top$. {The image (or column space) of a matrix $A$ is denoted as $\mr{image}(A)$, while its kernel (or null space) is denoted as $\mr{kernel}(A)$.} Consider a signal $w: \mb{Z}\to\mb{R}^{\dnw}$. The value of a signal $w:\mb{Z}\to\mb{R}^{\dnw}$ at time step~$k$ is denoted as $w(k)\in\mb{R}^{\dnw}$ and its $i$\tss{th} element %
is given by $w_i(k)\in\mb{R}$. The forward and backward time-shift operators are denoted by $\q$ and $\q^{-1}$. We denote a time-interval between $t_1$ and $t_2$, $t_1\leq t_2$ by $[t_1,t_2]\subset\mb{Z}$. For the time interval $\mb{T}\subseteq\mb{Z}$, we write $w_\mb{T}$ as the truncation to $w$ on $\mb{T}$, e.g., for $\mb{T}:=[1,N]$ we have $w_{[1,N]}=(w(1),\dots,w(N))\in\left(\mb{R}^\dnw\right)^{[1,N]}$. {The notation $w_{[1,N_1]}\land v_{[1,N_2]}\in\left(\mb{R}^\dnw\right)^{[1,N_1+N_2]}$ indicates concatenation of  $w_{[1,N_1]}$ followed by $v_{[1,N_2]}\in\left(\mb{R}^\dnw\right)^{[1,N_2]}$}, while, with a slight abuse of notation, $\col(w, v)$ indicates the stacked signal $(\dots, \begin{bsmallmatrix} w(k-1) \\ v(k-1) \end{bsmallmatrix}, \begin{bsmallmatrix} w(k) \\ v(k) \end{bsmallmatrix}, \begin{bsmallmatrix} w(k+1) \\ v(k+1) \end{bsmallmatrix}, \dots)$. A sequence of the following form $(p(k)\kron w(k))_{k=1}^{N}$ is denoted by $w^\mt{p}_{[1,N]}$. For $w_{[1,N]}$, the associated Hankel matrix of depth~$L$ is given by
\[ 
    \mc{H}_L(w_{[1,N]})=\begin{bmatrix}w(1) & w(2) & \dots & w({N-L+1})\\ w(2) & w(3) & \dots & w({N-L+2})\\ \vdots & \vdots & \ddots & \vdots \\ w({L}) & w({L+1}) & \dots & w({N})
\end{bmatrix},
\]
while the block-diagonal Kronecker operator `$\bkron$' is denoted as ${w}_{[1,N]}\bkron I_n =\mathrm{blkdiag}\big(w(1)\kron I_n, \dots, w(N)\kron I_n\big)$. {Finally, throughout the paper we distinguish signals from {recorded} data sets with a breve accent, e.g., $\breve{w}$.}

\section{Problem formulation}\label{s:prob}
\subsection{System definition and behaviors}
We study DT LPV systems that can be represented by the \emph{kernel} representation:
\begin{subequations}\label{eq:lpvker-shiftedaff}
\begin{equation}
    \underbrace{{\textstyle\sum_{i=0}^{\dnr}}r_i(\q^i p)\q^i}_{R(p,\q)} w = 0,
\end{equation}
with manifest signals $w\in\left(\mb{R}^{\dnw}\right)^\mb{Z}$, scheduling signals $p\in\mb{P}^\mb{Z}$ and scheduling dependent coefficients $r_{i}:\mb{P}^\mb{Z}\to\mb{R}^{\dnk\times\dnw}$ that have a \emph{shifted-affine} dependence on $p$:
\begin{equation}\label{eq:shifted-affine}
    r_i(\q^i p) = r_{i,0}+{\textstyle\sum_{j=1}^{\dnp}} r_{i,j}\q^i p_j,
\end{equation}
{with $r_{i,j}\in\mb{R}^{\dnk\times\dnw}$.} %
{We} refer to LPV systems that have a representation in the form of~\eqref{eq:lpvker-shiftedaff} as \emph{LPV Shifted-Affine} (LPV-SA) systems.
\end{subequations}
The signal $p$ is {considered to be free and} varying in the \emph{scheduling set} $\mb{P}\subseteq\mb{R}^{\dnp}$, which is often {chosen as} a closed subset of $\mb{R}^{\dnp}$ that contains the origin. {By introducing a partitioning\footnote{In many works on the behavioral approach, a non-singular permutation matrix $\Pi$ is used to characterize the partitioning, such that $w=\Pi\begin{bsmallmatrix} u \\ y \end{bsmallmatrix}$. To streamline the notation, we choose w.l.o.g. $\Pi=I$.} of $w$ to inputs $u\in(\mb{R}^\dnu)^\mb{Z}$ (maximally free\footnote{A maximally free input means that for a given $u$, none of the components of $y$ can be chosen freely for {all} $p\in\mf{B}_\mb{P}$.} signals) and outputs $y\in(\mb{R}^\dny)^\mb{Z}$, \eqref{eq:shifted-affine} becomes an} 
\emph{input-output} (IO) representation {with} $w=\col(u,y)$ and $r_i = [ \,r_{\mr{u},i} \  r_{\mr{y},i}\,]$, {{where} $r_{\mr{u},i}:\mb{P}^\mb{Z}\to\mb{R}^{\dnk\times\dnu}$, $r_{\mr{y},i}:\mb{P}^\mb{Z}\to\mb{R}^{\dnk\times\dny}$ and $\dnw=\dnu+\dny$.} %
{For} the remainder of this paper, the class of LPV-SA systems with $\dnp$ scheduling signals and $\dnw$ manifest variables {is denoted} by $\Sigma_{\dnp,\dnw}$.

In this paper, we consider the \emph{behavioral approach}~\cite{PoldermanWillems1997}. The behavior $\mf{B}$ is the collection of all solution trajectories compatible with the system. The representation~\eqref{eq:lpvker-shiftedaff} is a representation of a given behavior $\mf{B}$ of an LPV-SA system $\Sigma\in\Sigma_{\dnp,\dnw}$ if 
\begin{equation}\label{eq:shiftedaffinebehavior}
    \mf{B}=\{(w,p)\in(\mb{R}^{\dnw}\times\mb{P})^\mb{Z}\mid\text{\eqref{eq:lpvker-shiftedaff} holds}\}.
\end{equation}
{Following~\cite[Def.~3.24~\&~Def.~3.25]{Toth2010},} we consider~\eqref{eq:lpvker-shiftedaff} to be minimal if {$\dnk$ is such that $R(p,\q)$ in~\eqref{eq:lpvker-shiftedaff} has full row rank, and this full row rank representation has the smallest polynomial order $\dnr$ among all possible full row rank} kernel representations that can characterize $\mf{B}$ of a $\Sigma\in\Sigma_{\dnp,\dnw}$. {It is shown in~\cite[Thm.~3.6]{Toth2010} that there always exists a minimal kernel representations that represents $\mf{B}$.}

A few subsets of $\mf{B}$ are useful to consider. Specifically, the set of admissible scheduling trajectories of $\mf{B}$:
\begin{equation}\label{eq:BP}
    \mf{B}_\mb{P}  = \pi_\mr{p}\mf{B} = \{ p\in\mb{P}^\mb{Z} \mid \exists w\in\left(\mb{R}^{\dnw}\right)^\mb{T} \text{ s.t. }(w,p)\in\mf{B}\},
\end{equation}
the set of $w$ trajectories that are compatible with a {particular,} given
scheduling trajectory $p\in\mf{B}_\mb{P}$, i.e.,
\begin{equation}\label{eq:Bp}
    \mf{B}_p  = \{ w\in(\mb{R}^{\dnw})^\mb{T} \mid (w,p)\in\mf{B}\},
\end{equation}
and the trajectories in $\mf{B}$ that are restricted to the time interval $[k_1,k_2]\subset\mb{Z}$, $k_1\leq k_2$. The set containing these trajectories is given by
\begin{multline*}
    \Bfint{}{[k_1,k_2]}= \big\{(w,p)_{[k_1,k_2]}\in(\mb{R}^{\dnw}\times\mb{P})^{[k_1,k_2]}\,\big|\\ \exists \,(\omega,\rho)\in\mf{B}\text{ s.t. }(w(k),p(k))=(\omega(k),\rho(k)) \\ \text{for } k_1\le k\le k_2\big\}.
\end{multline*}
Note that this notation can be applied to~\eqref{eq:BP} and~\eqref{eq:Bp} as well. 

Because LPV systems are linear along a scheduling trajectory, $\mf{B}_p$ is a linear subspace. LPV systems are time invariant in the sense that $\q\mf{B}=\mf{B}$, and thus $\q\mf{B}_p=\mf{B}_{\q p}$. Moreover, $\Bfint{}{[k_1,k_2]}=\Bfint{}{[1, k_2-k_1+1]}$.

\begin{example}\label{exmp:msd}
    {We will clarify the introduced {concepts and} notation by means of {an} example.} We consider a \emph{mass-spring-damper} (MSD) system where the {stiffness of the} spring %
    is  varying {with} %
    a measurable signal $p$.
\begin{figure}
    \centering
    \includegraphics[width=0.65\linewidth]{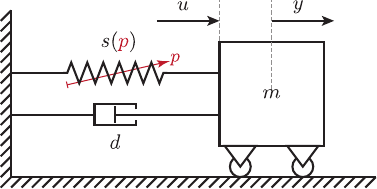}
    \caption{Schematics %
    of a mass-spring-damper system with a spring that is varying with a measurable signal~$p$. The input of the system is a force $u$, while the output is the measured position $y$.}
    \label{fig:msd}
\end{figure}
{Such an MSD system can occur in many practical scenarios,}
e.g., the linear axis of a water-jet cutter where the stiffness of the cable-slab is dependent on the water-flow of the jet, {{hence} $p$ {is} the (measurable) water-flow.} A schematic representation of the MSD system is given in Fig.~\ref{fig:msd}. The input $u$ is the force [N] exerted on the mass $m$ {[kg]}, and the output $y$ is the measured position [m] of the mass. The damper has damping coefficient~$d$, while the spring has a stiffness coefficient described by the function~$\springconstant(p)$. The variation of $\springconstant(p)$ is described by
\[ \springconstant(p(k)) = \springconstant_0+\springconstant_1 p(k), \quad p(k)\in[-1,1]. \]
Euler discretization of the CT MSD dynamics with sampling time $T_\mr{s}$ gives the following LPV representation:
\begin{multline}\label{eq:msd}
    y(k) +\left(\tfrac{d\,T_\mr{s}}{m}-2\right)y(k-1) \\ + \left(1+\tfrac{\springconstant_0\,T_\mr{s}^2-d\,T_\mr{s}}{m} + \tfrac{\springconstant_1\,T_\mr{s}^2}{m}p(k-2)\right)y(k-2) \\
    = \tfrac{T_\mr{s}^2}{m}u(k-2).
\end{multline}
{With $w=\mr{col}(u,y)$, we can write~\eqref{eq:msd} as an LPV-SA kernel representation of the form~\eqref{eq:lpvker-shiftedaff}, with
\[ R(p,\q) = \begin{bmatrix} -\tfrac{T_\mr{s}^2}{m} & \q^2 + \tfrac{d\,T_\mr{s}-2m}{m}\q + \tfrac{m+\springconstant_0\,T_\mr{s}^2-d\,T_\mr{s}}{m} + \tfrac{\springconstant_1\,T_\mr{s}^2}{m}p \end{bmatrix},\]
where we first multiplied both sides in~\eqref{eq:msd} with $\q^2$.
The solution trajectories for this system, collected in $\mf{B}$ are described by this particular $R(p,\q)$.}
\end{example}

\subsection{Problem statement}
The goal of this paper is to characterize $\Bfint{}{[1,L]}$ purely based on a given data set (Contribution~\ref{C:rep}). It is important that the characterization is \emph{computable} and %
it is verifiable %
whether the full $\Bfint{}{[1,L]}$ is represented by the {given} data (Contribution~\ref{C:cond}). Once we have these, we can use the representation for data-driven simulation (Contribution~\ref{C:sim}). This gives the following problem formulations:
\begin{problem}\label{problem:rep}
    Given a data set from an LPV-SA system $\breve{\Sigma}\in\Sigma_{\dnp,\dnw}$ with behavior $\mf{B}$ and a given complexity:
\begin{equation}
    \dataset= (\breve{w}_{[1,\Nd]}, \breve{p}_{[1,\Nd]})\in\Bfint{}{[1,\Nd]},
\end{equation}
where $\breve{w}$ and $\breve{p}$ are noise free. {For a given}~$L$, {formulate a representation of $\Bfint{}{[1,L]}$ of~$\breve\Sigma$ from only the data in $\dataset$}.
\end{problem}
{To ensure validity of the representation}, we require {computationally verifiable} conditions %
to conclude whether $\dataset$ is `rich' enough to represent $\Bfint{}{[1,L]}$:
\begin{problem}\label{problem:cond}
    Give an explicit condition for $\dataset$ that is {computationally} verifiable {to} %
     conclude whether $\Bfint{}{[1,L]}$ of $\breve{\Sigma}$ can be fully represented by $\dataset$.
\end{problem}
As motivated in Section~\ref{s:int}, the data-driven simulation problem for LPV systems was not formally solved. In this paper, we fill this gap by solving the following problem:
\begin{problem}\label{problem:sim}
    Use the solutions to Problems~\ref{problem:rep} and~\ref{problem:cond} to achieve simulation of $\breve{\Sigma}$, {i.e., computation of its response for a given input trajectory and initial condition, purely} based on $\dataset$.
\end{problem}

To obtain the solutions to the above problems, we first need to study the properties of LPV-SA systems and their (restricted) behaviors.

\section{Properties of LPV-SA behaviors}\label{s:sap}
In this section, we {explore} the properties of LPV-SA behaviors that are instrumental to solve Problems~\ref{problem:rep}--\ref{problem:sim}. We first provide the connections between kernel, IO and SS representations of LPV-SA behaviors. These connections naturally lead to the formal notion of complexity for an LPV-SA system, with which we formulate one of the key ingredients required to solve Problem~\ref{problem:cond}; the dimension of $\Bfint{p}{[1,L]}$. 

\subsection{Representations of LPV-SA behaviors}\label{ss:reps}
As motivated in Section~\ref{s:int}, the class of LPV-SA systems is, although restrictive, a highly useful system class. This is because it has direct IO and SS realizations with a structured dependence, which streamlines LPV identification, analysis and control design.
\subsubsection{Input-output representations}
A key aspect of the behavioral framework is that there is no prior distinction between inputs and outputs. This makes that kernel representations, e.g.,~\eqref{eq:lpvker-shiftedaff}, are the fundamental building blocks for representing systems in this framework. In control engineering, however, defining the input and output properties of signals is often needed. {Hence, following the partitioning $w=\col(u,y)$, as already introduced in Section \ref{s:prob}, we can represent the behavior in terms of an LPV-IO form.}

If the input is chosen such that it is {maximally free}, %
{then}
the number of inputs $\dnu$ is an invariant property of $\mf{B}$, {which follows directly from~\cite[Thm.~6]{willems1986time} and the definition of a maximally free input.} We denote the invariant property of the input dimension by $\mBf$, and this is, in fact, the first component of the \emph{complexity} of~$\mf{B}$. 

{We will consider an LPV-IO representation {of the behavior} in \emph{filter form}, {resulting from the transformation} $\q^{-\dnr}R(p,\q)w=0$. In this filter form, we can split up the coefficients of~\eqref{eq:lpvker-shiftedaff} according to the aforementioned partitioning, providing the $p$-dependent coefficient functions %
\[ r_i(\q^i p) = \begin{bmatrix} b_{\dnr-i}(\q^{i-\dnr}p) & -a_{\dnr-i}(\q^{i-\dnr}p) \end{bmatrix}, \quad i\in\mb{I}_0^\dnr, \]
with}
\begin{equation}\label{eq:abpoly}
    a(p,\q)  = \sum_{i=0}^{\dna}a_{i}(\q^{-i}p)\q^{-i}, \quad
    b(p,\q)  = \sum_{i=0}^{\dnb}b_{i}(\q^{-i}p)\q^{-i},
\end{equation}
and $\dnr=\max(\dna,\dnb)$, {giving} the shifted-affine LPV-IO realization
\begin{equation}\label{eq:lpvio-shifted}
    y(k) + \sum_{i=1}^{\dna}a_i(p(k-i)) y(k-i) = \sum_{j=0}^{\dnb}b_j(p(k-j)) u(k-j).
\end{equation}
{Here, we assume that $a_0=I_\dny$. %
Although for \emph{single-input-single-output} (SISO) systems, dividing by $a_0$ is always possible to obtain~\eqref{eq:lpvio-shifted}, {such an operation} can potentially introduce rational dependency over $p(k)$ in the remaining coefficients with corresponding singularities. 
In the \emph{multiple-input-multiple-output}~(MIMO) case, invertibility of $a_0$ is not always ensured {to obtain such a form}. {Nevertheless, in} the sequel, we restrict our attention to cases where the leading coefficient $a_0$ can be taken to be identity, for the purpose of showing the connections between the different representations.}

With $a_0=I_\dny$, the scheduling-dependent functions $a_i,b_i$ in~\eqref{eq:lpvio-shifted} are simply partitions of the coefficient functions $r_i$ in~\eqref{eq:lpvker-shiftedaff}, hence they are affine in $p(k-i)$, {and $u$ is a maximally free signal w.r.t. each $\mf{B}_p$.}
Minimality of the LPV-IO representation~\eqref{eq:lpvio-shifted} is directly adopted from the kernel representation, and is achieved if~$a$ in~\eqref{eq:abpoly} has full row rank. The LPV-IO representation is controllable if it is minimal and the polynomials $a$ and $b$ are \emph{left-coprime}, {see Def.~3.28 and the proof of Thm.~5.1 in~\cite{Toth2010}}. The latter implies that~$a$ and $b$ contain the minimum number of lags of $u$ and $y$ to represent~$\mf{B}$.  Here we uncover another invariant property of~$\mf{B}$; the \emph{lag} $\LBf$, which is the minimum required degree of the polynomials in~\eqref{eq:abpoly} to be able to represent $\mf{B}$. Hence, for a minimal~\eqref{eq:lpvio-shifted}, $\max(\dna,\dnb) = \dnr = \LBf$. The lag is another component of the complexity of $\mf{B}$. The final measure of the complexity of $\mf{B}$ is the minimal required state dimension of an LPV-SS {realization of $\mf{B}$}, which we will discuss next.

\subsubsection{State-space representations}\label{sss:lpvsses}
LPV-SS representations are standard in LPV analysis and control design. Particularly useful are LPV-SS representations with \emph{static} scheduling dependence, i.e., {composed of matrices that} are only dependent on $p(k)$:
\begin{subequations}\label{eq:lpvssgen}
    \begin{align}
        \q x & = A(p) x + B(p) u, \\
        y & = C(p) x + D(p) u,
    \end{align}
\end{subequations}
with $x(k)\in\mb{R}^{\dnx}$ being {a latent variable that qualifies as a} state and  $u(k)\in\mb{R}^{\mBf}$. The \emph{full behavior} of~\eqref{eq:lpvssgen} is given by
\begin{multline}
    \Bss = \{(u,y,p,x)\in\left(\mb{R}^\dnu\times\mb{R}^\dny\times\mb{P}\times\mb{R}^\dnx\right)^\mb{Z} \mid \\ \text{\eqref{eq:lpvssgen} holds}\}.
\end{multline}
An important feature of LPV-SA systems is that the LPV-IO representation~\eqref{eq:lpvio-shifted} has a \emph{direct} LPV-SS realization of the form~\eqref{eq:lpvssgen}~\cite{toth2011state} with matrix functions %
\setlength{\dashlinegap}{2pt}
\begin{multline}\label{eq:lpvss}
    \left[\begin{array}{c:c} A(p) & B(p) \\\hdashline C(p) & D(p)\end{array} \right] = \\
    \left[\begin{array}{cccc:c} 
            \!\!\!\unaryminus a_{1}(p) \!&\! I_\dny   & \cdots & 0 \ & b_{1}(p)-a_1(p)b_0(p)\\
            \vdots & \vdots & \ddots  & \vdots & \vdots \\
            \!\!\!\unaryminus a_{n_\mr{a}\!\unaryminus 1}(p) \!&\! 0 & \cdots  & I_\dny & b_{n_\mr{b}\!\unaryminus 1}(p)\!-\!a_{n_\mr{b}\!\unaryminus 1}(p)b_0(p)\!\!\!\\
            \!\unaryminus a_{n_\mr{a}}(p) \!&\! 0 &  \cdots & 0 & b_{n_\mr{b}}(p)-a_{n_\mr{b}}(p)b_0(p) \\\hdashline 
            I_\dny & 0 & \cdots & 0 & b_0(p)
    \end{array} \right],
\end{multline}
where, with a slight abuse of notation, the state construction is,  {see~\cite[Sec.~IV.A]{toth2011state},}
\begin{subequations}\label{eq:stateconstruction}
\begin{equation}
    x_i = \q x_{i-1} + a_{i-1}(p)y-b_{i-1}(p)u, \quad  x_i(k)\in\mb{R}^{\dny}
\end{equation}
for $i\in[2, \max(\dna,\dnb)]$, and where
\begin{equation}\label{eq:stateconstruction:x1}
    x_1 = y - b_0(p)u.
\end{equation}
\end{subequations}
Note that with the construction~\eqref{eq:stateconstruction}, $\dnx=\dny\LBf$.

Due to the direct LPV-SS realization, we have that $\pi_\mr{u,y,p}\Bss=\mf{B}$, i.e., the manifest behaviors defined by the IO representation~\eqref{eq:lpvio-shifted} and the LPV-SS representation~\eqref{eq:lpvssgen} with~\eqref{eq:lpvss} are equivalent. {Based on this}, we can also define minimality of the LPV-SS representation~\eqref{eq:lpvssgen} {(with static scheduling dependence)} as the minimum number of states required such that $\pi_\mr{u,y,p}\Bss=\mf{B}$ holds {within the class $\Sigma_{\dnp,\dnw}$}. We call this the \emph{order} of $\mf{B}$. This is an invariant property for {behaviors $\mf{B}$ of LPV-SA systems in $\Sigma_{\dnp,\dnw}$, which follows from~\cite[Thm.~3.7]{Toth2010}}. This, in fact, is the last measure for the complexity of an LPV-SA system, which we denote by~$\nBf$. 

For SISO systems, i.e., $\dnu=\dny=1$, the direct LPV-SS realization~\eqref{eq:lpvss} is minimal, i.e., $\dnx=\nBf$, if the polynomials~$a$ and~$b$ in~\eqref{eq:lpvio-shifted} are left-coprime. This means that we also have $\LBf=\nBf$. For MIMO systems, i.e., $\dnu,\dny>1$, this is generally not the case {as} %
$\LBf\leq\nBf$. In the MIMO case, a minimal realization of~\eqref{eq:lpvssgen} {from}~\eqref{eq:lpvss} can always be obtained by means of moment matching\footnote{See also the implementation in \lpvcore~\cite{BoefCoxToth2021}.}~\cite{BastigPetreczkyToth2015} or an LPV Kalman decomposition~\cite{petreczky2016realization}. With both methods, the minimal realization of~\eqref{eq:lpvssgen} is obtained using a \emph{constant} projection matrix that projects the state to a lower dimension. This means that the resulting (reduced) $A(p),\dots,D(p)$ will still have static scheduling dependence and the \emph{same} functional dependence (e.g., $C$ is still scheduling independent). 
 
 \subsection{Complexity and dimension of behaviors}
From Section~\ref{ss:reps}, we recovered the integers $\mBf$, $\LBf$, and $\nBf$ that are a measure for the complexity of the behavior $\mf{B}$ of an LPV-SA system $\Sigma\in\Sigma_{\dnp,\dnw}$. In line with~\cite{markovskyVerhoek2024}, we characterize the complexity by the triple
\begin{equation}\label{eq:complexity}
    \cBf = (\mBf, \LBf, \nBf),
\end{equation}
where $\mBf$ is the number of inputs, $\LBf$ is the minimal lag of the system, and $\nBf$ is the order of $\mf{B}$. 

With these integer invariants defined, we will now formulate one of the key ingredients required for the solution to Problem~\ref{problem:cond}. More specifically, we now show that for an $L\geq\LBf$, the dimension of $\Bfint{p}{[1,L]}$ with $p_{[1,L]}\in\Bfint{\mb{P}}{[1,L]}$ is equal to $\nBf +\mBf L$.
\begin{lemma}[Dimension of $\Bfint{p}{[1,L]}$]\label{lem:dim}
    Consider an LPV-SA system $\Sigma\in\Sigma_{\dnp,\dnw}$ with behavior $\mf{B}$ and complexity $\cBf$. Given any $p_{[1,L]}\in\Bfint{\mb{P}}{[1,L]}$. Then, $\dim(\Bfint{p}{[1,L]}) = \nBf +\mBf L$ if and only if $L\geq\LBf$.
\end{lemma}
\begin{proof}
    See Appendix~\ref{app:pf:lemdim}.
\end{proof}
This result will allow us to prove what we call the LPV Fundamental Lemma for the class of LPV-SA systems~$\Sigma_{\dnp,\dnw}$. Note that we have presented a version of Lemma~\ref{lem:dim} in the meromorphic context in {preliminary work}~\cite[Cor.~1]{Verhoek2021_CDC}. In this paper, we prove this result in the context of the class of LPV-SA systems.

We now have all the ingredients for the formulation of a data-driven representation of the finite-horizon behavior of LPV-SA systems, and thus {to} solve Problems~\ref{problem:rep} and~\ref{problem:cond}.

\section{Data-driven representation of LPV-SA systems}\label{s:main}
This section presents the first part of our main result, which is the data-driven characterization of LPV-SA systems. {First, a representation of} the finite-horizon behavior {of LPV-SA systems based on a} %
given data set~$\dataset$ is introduced, providing a solution to Problem~\ref{problem:rep}. {Next, we derive a GPE condition to test}
whether~$\dataset$ is `rich' enough %
{to}
fully characterize the finite-horizon behavior, providing {a} solution to Problem~\ref{problem:cond}. 
We conclude this section with a note on input design.

\subsection{Data-driven representation}\label{sec:shiftedaff_ddrep}
\begin{figure}
    \centering
    \begin{subfigure}[b]{0.325\linewidth}
        \centering
        \includegraphics[width=0.8\textwidth]{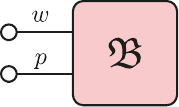}
        \caption{\vspace{-5pt}}
    \end{subfigure}~\begin{subfigure}[b]{0.325\linewidth}
        \centering
        \includegraphics[width=0.8\textwidth]{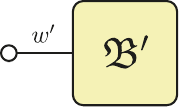}
        \caption{\vspace{-5pt}}
    \end{subfigure}~\begin{subfigure}[b]{0.325\linewidth}
        \centering
        \includegraphics[width=0.8\textwidth]{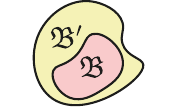}
        \caption{\vspace{-5pt}}\label{fig:ltiembedding:c}
    \end{subfigure}
    \caption{{Illustration of the (a) LPV system versus (b) its LTI embedding and (c) their corresponding behaviors. The behavior of the LTI embedding over-approximates the behavior of the LPV system.}
    }\label{fig:ltiembedding}
\end{figure}

We formulate a data-driven representation of $\Bfint{p}{[1,L]}$ that is valid for any $p\in\Bfint{\mb{P}}{[1,L]}$ by means of embedding {the behavior represented by} the kernel representation into an LTI realization, whose behavior is constrained by a scheduling-dependent kernel constraint. Isolating a single term in the polynomial corresponding to the kernel representation~\eqref{eq:lpvker-shiftedaff}
\begin{equation}\label{eq:lpvker-sa-single}
    \left(r_{i,0} + {\sum_{j=1}^{\dnp}}r_{i,j}p_j(k+i)\right)w(k+i),
\end{equation}
{reveals that we can write the individual terms of~\eqref{eq:lpvker-shiftedaff}, i.e.,~\eqref{eq:lpvker-sa-single},} in terms of the \emph{auxiliary signal}
\begin{equation}\label{eq:wprime}
    w^\prime(k+i)= \begin{bmatrix} 1 \\ p(k+i) \end{bmatrix}\kron w(k+i)\in\mb{R}^{(1+\dnp)\dnw},
\end{equation}
such that 
\begin{multline}
        \left(r_{i,0} + {\textstyle\sum_{j=1}^{\dnp}}r_{i,j}p_j(k+i)\right)w(k+i) = \\ \underbrace{\begin{bmatrix} r_{i,0} & \cdots & r_{i,\dnp} \end{bmatrix}}_{r_i^\prime} w^\prime(k+i).
\end{multline}
The definition of this auxiliary signal $w^\prime$ allows us to embed the behavior associated with a kernel representation with shifted-affine dependency as an \emph{LTI representation}. Intuitively, this LTI embedding {treats} the $\dnp\dnw$ variables of $p(k)\kron w(k)$ %
as additional inputs, i.e., free variables, {even if they are not}.  The LTI embedding of~\eqref{eq:lpvker-shiftedaff} gives
\begin{equation}\label{eq:LTIembedding}
    R^\prime(\q) w^\prime = {\textstyle\sum_{i=0}^{\dnr}}r^\prime_i \q^i w^\prime = 0, 
\end{equation}
with the behavior
\begin{equation}\label{eq:BehLTIembedding}
    \mf{B}^\prime=\{w^\prime : \mb{T}\to\mb{R}^{(1+\dnp)\dnw} \mid \text{\eqref{eq:LTIembedding} holds}\}.
\end{equation}
{The main difference between} $\mf{B}$ and $\mf{B}^\prime$ is that the entries of $w^\prime$ are \emph{not} independent from each other in the original LPV representation~\eqref{eq:lpvker-shiftedaff}, while in the LTI embedding~\eqref{eq:LTIembedding} this interdependency is \emph{ignored}. 
Hence, for some $w^\prime\in\mf{B}^\prime$ there might not exists a pair $(w,p)\in\mf{B}$ such that $w^\prime(k)=\col(w(k), p(k)\kron w(k))$ for all $k$, implying that $\mf{B}\subset\mf{B}^\prime$. The resulting behavior of the LTI embedding thus \emph{over-approximates} the behavior of the LPV system, see Fig.~\ref{fig:ltiembedding:c} for illustration. {We will counteract this over-approximation in the next paragraph, to arrive at an exact data-driven LPV representation of $\Bfint{p}{[1,L]}$.}
As the introduced auxiliary signals $p(k)\kron w(k)\in\mb{R}^{\dnp\dnw}$ are considered to be free, we have that 
\begin{equation}\label{eq:mBfprime}
    \mBpf = \mBf+\dnp\dnw.
\end{equation}
Additionally, this consideration {implies} that~\eqref{eq:LTIembedding} is an LTI system on which we can apply the existing results on data-driven LTI representations, i.e., Willems' Fundamental Lemma and its associated results~\cite{markovsky2022identifiability, WillemsRapisardaMarkovskyMoor2005}. {For a given data set $\dataset^\prime=\breve{w}^\prime_{[1,\Nd]}$, these results provide that} %
\begin{equation}\label{eq:FLltiemb}
    \mr{image}(\mc{H}_L(\breve w^\prime_{[1,\Nd]})) \subseteq \left.\mf{B}^\prime\right|_{[1,L]},
\end{equation}
where equality holds if and only if the GPE (\emph{generalized persistence of excitation}) condition for LTI systems~\cite[{Cor.~19}]{markovsky2022identifiability} holds, i.e.,
\begin{equation}\label{eq:GPE_ltiembedding}
    \mr{rank}(\mc{H}_L(\breve w^\prime_{[1,\Nd]})) = \nBf + \mBpf L.
\end{equation}
Given that~\eqref{eq:GPE_ltiembedding} holds, we thus have that for any $w^\prime_{[1,L]}\in\left.\mf{B}^\prime\right|_{[1,L]}$, there exists a $g\in\mb{R}^{\Nd-L+1}$ such that 
\begin{equation}\label{eq:fl_ltiembed}
    \mc{H}_L(\breve w^\prime_{[1,\Nd]})g = \mr{vec}(w^\prime_{[1,L]})
\end{equation}
holds. Hence, the equality~\eqref{eq:fl_ltiembed} serves as a data-driven representation of the extended behavior $\mf{B}^\prime$ of the LTI embedding. We will now take back into account the previously ignored interdependencies in $w^\prime$, {and thus counteract the aforementioned over-approximation}.

Let us introduce the signal 
\( w^\mt{p}_{[1,\Nd]} = (p(k)\kron w(k))_{k=1}^{\Nd}\)
such that
\( w^\prime_{[1,\Nd]}=\col\big(w_{[1,\Nd]}, w^\mt{p}_{[1,\Nd]}\big). \)
This allows us to write~\eqref{eq:fl_ltiembed}, after a permutation {of} the rows, as 
\begin{equation}\label{eq:fl_lti_writtenout}
    \begin{bmatrix} \vphantom{\Big)}\mc{H}_L(\breve w_{[1,\Nd]}) \\ \mc{H}_L(\breve{w}^{\breve{\mt{p}}}_{[1,\Nd]}) \end{bmatrix}g=\begin{bmatrix} \vphantom{\Big)} \mr{vec}(w_{[1,L]}) \\ \mr{vec}(w^\mt{p}_{[1,L]}) \end{bmatrix}.
\end{equation}
As observed in~\cite{VerhoekAbbasTothHaesaert2021}, $\mr{vec}(w^\mt{p}_{[1,L]}) = \mc{P}^{\dnw}\mr{vec}(w_{[1,L]})$, where %
$\mc{P}^{\dnw}= p_{[1,L]}\bkron I_{\dnw}$. Hence,
\begin{equation}\label{eq:restrictionRHS}
    \mr{vec}(w^\mt{p}_{[1,L]}) = \mc{P}^{\dnw}\mr{vec}(w_{[1,L]}) = \mc{P}^{\dnw}\mc{H}_L(\breve w_{[1,\Nd]})g.
\end{equation}
Any $g$ satisfying~\eqref{eq:restrictionRHS}, respects the underlying dynamic structure w.r.t. the scheduling in $w^\mt{p}_{[1,L]}$, i.e., it serves as a \emph{restriction} on the behavior of the LTI embedding {that {\emph{eliminates}} the aforementioned over-approximation}. {This restriction becomes visible when we substitute~\eqref{eq:restrictionRHS} into~\eqref{eq:fl_lti_writtenout}, and move~`$\mc{P}^{\dnw}\mc{H}_L(\breve w_{[1,\Nd]})g$' to the left-hand side of~\eqref{eq:fl_lti_writtenout}:}
\begin{equation}\label{eq:fundamentallemmafull}
    \begin{bmatrix} \mc{H}_L(\breve w_{[1,\Nd]}) \\ \mc{H}_L(\breve{w}^{\breve{\mt{p}}}_{[1,\Nd]})-\mc{P}^\dnw \mc{H}_L(\breve w_{[1,\Nd]})\end{bmatrix}g=\begin{bmatrix} \mr{vec}(w_{[1,L]}) \\ 0 \end{bmatrix}.
\end{equation}
{We can notice}
that the first block-row in~\eqref{eq:fundamentallemmafull} characterizes the LTI part of the LPV system (associated with $r_{i,0}$), while the second block-row provides a restriction on the $g$ vectors that provide {trajectories} $w_{[1,L]}$ from the linear combination of the columns of $\mc{H}_L(\breve w_{[1,\Nd]})$. This restriction is not only dependent on the information encoded in $(\breve{w}_{[1,\Nd]}, \breve{p}_{[1,\Nd]})$, i.e., $\dataset$, but also on the scheduling signal $p_{[1,L]}\in\Bfint{\mb{P}}{[1,L]}$ associated with $w_{[1,L]}$. {Hence, the left-hand side of~\eqref{eq:fundamentallemmafull} provides us with a \emph{data-driven} characterization of $\Bfint{p}{[1,L]}$, i.e., a characterization of all $w_{[1,L]}$ trajectories that correspond to the scheduling trajectory $p_{[1,L]}\in\Bfint{\mb{P}}{[1,L]}$.} 

Given sufficiently rich data, we now established that all possible $w_{[1,L]}\in\Bfint{p}{[1,L]}$ are characterized by vectors $g$  that satisfy~\eqref{eq:fundamentallemmafull}, i.e., vectors $g$ that are both in the row space of $\mc{H}_L(\breve w_{[1,\Nd]})$ and the kernel of $\mc{H}_L(\breve{w}^{\breve{\mt{p}}}_{[1,\Nd]})-\mc{P}^\dnw \mc{H}_L(\breve w_{[1,\Nd]})$ for a given scheduling trajectory $p_{[1,L]}$. Hence, by defining
\begin{equation}\label{eq:defNp}
    \meu{N}_{p}= \mr{kernel}\left(\mc{H}_L(\breve{w}^{\breve{\mt{p}}}_{[1,\Nd]}) - \mc{P}^{\dnw}\mc{H}_L(\breve{w}_{[1,\Nd]}) \right),
\end{equation}
we have that
\begin{equation}\label{eq:imHsubseteqB}
    \mr{image}\left(\mc{H}_L(\breve{w}_{[1,\Nd]})\meu{N}_{p}\right)\subseteq\Bfint{p}{[1,L]}.
\end{equation}
We now established an \emph{exact} data-driven representation of the set of $w_{[1,L]}$ sequences associated with $p_{[1,L]}$, corresponding to Problem~\ref{problem:rep} (Contribution~\ref{C:rep}). Next, we will establish a condition on $\dataset$ that guarantees~\eqref{eq:imHsubseteqB} to hold with equality, providing the solution to Problem~\ref{problem:cond} (Contribution~\ref{C:cond}).

\subsection{The LPV fundamental lemma}
In this section, we provide a necessary and sufficient condition verifiable from the data $\dataset$ that ensures whether~\eqref{eq:imHsubseteqB} holds with equality, providing Contribution~\ref{C:cond}. We establish this by showing that the dimension of $\mr{image}\left(\mc{H}_L(\breve{w}_{[1,\Nd]}\meu{N}_{p}\right)$ is strongly linked to the dimensionality of $\Bfint{p}{[1,L]}$. This results in an so-called ``identifiability condition'' for LPV-SA systems, i.e., a GPE condition that is analogous to the condition put forward in~\cite{markovsky2022identifiability}. In other words, we prove the necessary and sufficient conditions that the data in $\dataset$ should satisfy to be able to characterize the full $\Bfint{{p}}{[1,L]}$ for a given $p_{[1,L]}\in\Bfint{\mb{P}}{[1,L]}$.
\begin{theorem}[LPV-SA Fundamental Lemma]\label{thm:FLeasy}
    Given a data set $\dataset\in\Bfint{}{[1,\Nd]}$ from an LPV-SA system $\breve{\Sigma}\in\Sigma_{\dnp,\dnw}$. {Construct} %
    $\meu{N}_p$ for some ${p}_{[1,L]}\in\Bfint{\mb{P}}{[1,L]}$ as in~\eqref{eq:defNp}. For $L\ge\LBf$, {the following statements are equivalent:
    \begin{enumerate}[label=\textit{\roman*).}, align=left, ref={(\roman*)}, leftmargin=*]
        \item \label{thmitem:beh} For all ${p}_{[1,L]}\in\Bfint{\mb{P}}{[1,L]}$,
            \begin{equation} \label{eq:thm:beh}
                \Bfint{p}{[1,L]} = \mr{image}\left(\mc{H}_L(\breve{w}_{[1,\Nd]})\meu{N}_{p}\right),
            \end{equation}
        \item \label{thmitem:dim} For all ${p}_{[1,L]}\in\Bfint{\mb{P}}{[1,L]}$,
            \begin{equation} \label{eq:thm:dim}
                \mr{rank}\left(\mc{H}_L(\breve{w}_{[1,\Nd]})\meu{N}_{p}\right)= \nBf + \mBf L,
            \end{equation}
        \item For any $({w}_{[1,L]},{p}_{[1,L]})\in\Bfint{}{[1,L]}$, there exists a vector $g\in\mb{R}^{\Nd-L+1}$ such that~\eqref{eq:fundamentallemmafull} holds, \label{thmitem:rep}
        \item The following rank condition is satisfied:
            \begin{equation}\label{eq:GPE}
            \mr{rank}\left(\begin{bmatrix} \vphantom{\Big)} \mc{H}_L(\breve w_{[1,\Nd]}) \\ \mc{H}_L(\breve{w}^{\breve{\mt{p}}}_{[1,\Nd]}) \end{bmatrix}\right) = \nBf + (\mBf+\dnp\dnw)L.
            \end{equation}\label{thmitem:gpe}
    \end{enumerate}}
\end{theorem}
\begin{proof}
See Appendix~\ref{app:pf:fl}.
\end{proof}
Item~\ref{thmitem:dim} in Theorem~\ref{thm:FLeasy} provides a necessary and sufficient dimensionality condition on the data-driven representation of~$\Bfint{p}{[1,L]}$ that is verifiable from the data set $\dataset$. Specifically, the data can fully represent $\Bfint{p}{[1,L]}$ for any $p_{[1,L]}\in\Bfint{\mb{P}}{[1,L]}$ if and only if~\eqref{eq:thm:dim} holds for all $p_{[1,L]}\in\Bfint{\mb{P}}{[1,L]}$. Although, %
\eqref{eq:thm:dim} seems to result in an infinite test over all possible $p_{[1,L]}\in\Bfint{\mb{P}}{[1,L]}$, through the LTI embedding, we show {with Item~\ref{thmitem:gpe}} that this reduces to a single, simple rank test on the left-hand side of~\eqref{eq:fl_lti_writtenout}, which is only composed from the given data in~$\dataset$.
We will refer to \eqref{eq:GPE} as the LPV-GPE condition.
The LPV-GPE condition also provides %
a lower bound for $\Nd$:
\begin{equation}\label{eq:minimumNd}
    \Nd \geq \big(1+\dnw\dnp+\mBf\big)L+\nBf-1,
\end{equation}
i.e., the minimum number of samples in $\dataset$ required to represent $\Bfint{p}{[1,L]}$ for an arbitrary $p_{[1,L]}\in\Bfint{\mb{P}}{[1,L]}$. {{It is important to highlight that the LPV-GPE condition provides no separate PE condition}  
on the {used} scheduling trajectory~{$\breve{p}_{[1,\Nd]}$ or input signal $\breve{u}_{[1,\Nd]}$} in the data set, only on the {joint} collection of {input, scheduling, and output} signals in the data-dictionary (through~\eqref{eq:GPE}). In the next section, we provide a discussion on generating a~$\dataset$ that satisfies~\eqref{eq:GPE}.}

\subsection{Input design}
What makes Theorem~\ref{thm:FLeasy} different from the original (LPV) Fundamental Lemma in~\cite{WillemsRapisardaMarkovskyMoor2005, Verhoek2021_CDC} is that we now have a rank condition on the Hankel matrices involving trajectories of $w$ and $p$, while in~\cite{WillemsRapisardaMarkovskyMoor2005}, the rank condition is only on the input signal $u$, specifically, $\rank(\mc{H}_{L+\nBf}(\breve{u}_{[1,\Nd]})) = \mBf(L+\nBf)$. The latter yields an \emph{input design} condition, which allows you to \emph{a priori} design an experiment for the construction of a data-driven representation, without taking the output of the system itself into account. It would be tempting to adopt this condition for the LPV-SA case, i.e., by taking this rank condition on the Hankel matrix of $\begin{bsmallmatrix} u \\ u^\mt{p}\end{bsmallmatrix}$. This would lead to the condition that if 
\begin{equation}\label{eq:wrongcondition}
    \rank\!\left(\!\begin{bmatrix} \mc{H}_{L+\nBf}(\breve{u}_{[1,\Nd]}) \\ \mc{H}_{L+\nBf}(\breve{u}^{\breve{\mt{p}}}_{[1,\Nd]}) \end{bmatrix}\!\right) = \mBf(1+\dnp)(\nBf+L),
\end{equation}
then \eqref{eq:thm:beh} holds. In the following counter example, we show that this is, unfortunately, not the case.
\begin{example}\label{exmp:counter}
    Consider an LPV-SA system $\Sigma\in\Sigma_{\dnp,\dnw}$ with the LPV-IO representation:
    \[ y(k) + (1 + p(k-1))y(k-1) = u(k) + p(k-1)u(k-1),  \]
    which has shifted-affine scheduling dependence. 
    Note that $\nBf=\mBf=\LBf=\dnp=1$ and $\dnw=2$ for this particular system.
    We compute a data-driven representation of~$\Sigma$ for $L=10$. For this, we generate a $\dataset$ with $\Nd=40$, according to \eqref{eq:minimumNd}. With $y(0)=u(0)=p(0)=1$, we apply {an i.i.d.} scheduling signal $p_{[1,\Nd]}$ with $p(k)\sim\mc{N}(0,1)$ and an input that is constructed as:
    \[ u(k) = p(k-1)(1-u(k-1)) + 2, \]
    to the system. Note that this input can in fact be seen as a control policy to regulate the system to $y(k)=1$ for any scheduling sequence. The resulting input and scheduling sequences are shown in Fig.~\ref{fig:counterex1}.
    \begin{figure}
        \centering
        \includegraphics[width=0.975\linewidth]{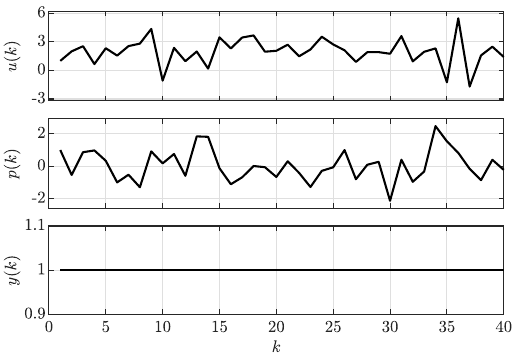}
        \caption{Input, scheduling, and output sequences {of} $\dataset$ {in} Example~\ref{exmp:counter}.}
        \label{fig:counterex1}
    \end{figure}
    By looking at these signals, one could already argue that they are persistently exciting, and, indeed, computing the rank in condition~\eqref{eq:wrongcondition} gives a rank of~$22$ (note that $\mBf(1+\dnp)(L+\nBf)=22$). %
    However, when inspecting the output response, we see that $y(k)=1$ for all $k=1,\dots,\Nd$, which is due to the construction of $u(k)$. This means that~\eqref{eq:thm:dim} and thus~\eqref{eq:thm:beh} will never be satisfied, showing that directly adopting the \emph{input}-design condition from~\cite{WillemsRapisardaMarkovskyMoor2005} does not apply for systems of the class $\Sigma_{\dnp,\dnw}$. Applying the condition on the LTI embedding~\eqref{eq:LTIembedding} with inputs $\col(u, w^\mt{p})$, gives the correct conclusion, i.e.,
    \begin{multline}\label{eq:pe-condition}
        \rank\left(\begin{bmatrix} \mc{H}_{L+\nBf}(\breve{u}_{[1,\Nd]}) \\ \mc{H}_{L+\nBf}(\breve{u}^{\breve{\mt{p}}}_{[1,\Nd]}) \\ \mc{H}_{L+\nBf}(\breve{y}^{\breve{\mt{p}}}_{[1,\Nd]}) \end{bmatrix}\right) \\ = \underbrace{(\mBf+\dnw\dnp)}_{\mBpf}(L+\nBf),
    \end{multline}
    should hold in order for~\eqref{eq:thm:beh} to hold. Computing the rank as in~\eqref{eq:pe-condition} with the obtained data set gives~$24$, while $(\mBf+\dnw\dnp)(L+\nBf)=33$, concluding that the data cannot represent the considered system for $L=10$.
\end{example}
From Example~\ref{exmp:counter}, we see that, in order to have a condition on the input (and scheduling) that a priori guarantees~\eqref{eq:thm:beh} to hold, we need a condition on the design of $(\breve u_{[1,\Nd]},\breve p_{[1,\Nd]})$ which guarantees that the resulting $(\col(\breve u_{[1,\Nd]},\breve y_{[1,\Nd]}),\breve p_{[1,\Nd]})\in\Bfint{}{[1,\Nd]}$ will satisfy~\eqref{eq:pe-condition}. We currently do not have a systematic solution to this problem. {Hence, we currently only have methods that can \emph{a posteriori} verify whether the data satisfies the LPV-GPE condition. The development of input (and scheduling) design conditions that \emph{a priori} guarantee satisfaction of~\eqref{eq:GPE} is} an important and interesting topic for future research.

\section{Data-driven simulation}\label{s:sim}

In this section, we consider the solution to the simulation problem in a data-driven setting {in terms of} Problem~\ref{problem:sim}. We want to emphasize that, although the data-driven simulation %
has been {used already} %
in the LPV setting, see~\cite{Verhoek2021_CDC, VerhoekAbbasTothHaesaert2021, verhoek2023linear}, it has never been worked out in detail from a theoretical perspective. Therefore, we {formally work out} the solution to the simulation problem in this section as a generalization of the LTI data-driven simulation that is presented in~\cite{MarkovskyRapisarda2008}.
\begin{figure}
    \centering
    \includegraphics[width=0.975\linewidth]{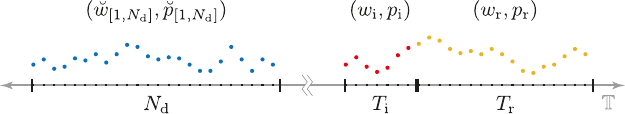}
    \caption{Schematic representation of the simulation problem: With a length-$T_\mr{i}$ initial trajectory $(w_\mr{i},p_\mr{i})$ (depicted by the red data points), determine the response $(w_\mr{r},p_\mr{r})$ (depicted by the yellow data points) using only a given data set $\dataset = (\breve{w}_{[1,\Nd]}, \breve{p}_{[1,\Nd]})$ (depicted by the blue data points).}
    \label{fig:simulation}
\end{figure}
\subsection{The LPV data-driven simulation problem}\label{ss:simprob}
The {general} %
form %
of the %
simulation problem is: given an LPV-SA system $\Sigma\in\Sigma_{\dnp,\dnw}$ with IO partition $w=\col(u,y)$,  {starting from an initial condition,} %
find the response~$y_\mr{r}$ of $\Sigma$ to $(u_\mr{r}, p_\mr{r})$ such that $(\col(u_\mr{r}, y_\mr{r}), p_\mr{r})\in\Bfint{}{[1,T_\mr{r}]}$. 

Classically, {the initial condition} is characterized by {an} initial state $\mt{x}$ of an LPV-SS representation of $\mf{B}$, {but equivalently such an} %
initial condition can be (uniquely) expressed in terms of an \emph{initial trajectory}. Hence, this provides a way {to} formalize the \emph{data-driven simulation problem}{, i.e., Problem~\ref{problem:sim},} as follows:
\begin{problem}\label{problem:simformal}
Given a data set $\dataset\in\Bfint{}{[1,\Nd]}$ from an LPV-SA system $\breve\Sigma\in\Sigma_{\dnp,\dnw}$, an input-scheduling pair $(u_\mr{r}, p_\mr{r})\in\left(\mb{R}^\dnu\times\mb{P}\right)^{[1,T_\mr{r}]}$ and an initial trajectory $(w_\mr{i}, p_\mr{i})\in\Bfint{}{[1,T_\mr{i}]}$. Find a response $y_\mr{r}$ of $\breve\Sigma$ to the input-scheduling pair $(u_\mr{r}, p_\mr{r})\in\left(\mb{R}^\dnu\times\mb{P}\right)^{[1,T_\mr{r}]}$, %
such that $(w_\mr{i},p_\mr{i})\land(\col(u_\mr{r}, y_\mr{r}), p_\mr{r})\in\Bfint{}{[1,T_\mr{i}+T_\mr{r}]}$. 
\end{problem}
The simulation problem is also illustrated in Fig.~\ref{fig:simulation}.

\subsection{Trajectory-based initial condition}

{First, we formally} show that the initial condition $\mt{x}$ of the response $(w_\mr{r},p_\mr{r})$ can be uniquely determined with a length-$T_\mr{i}$ initial trajectory if $T_\mr{i}$ is larger than $\LBf$. 
\begin{lemma}[Initial state characterization]\label{lem:state}
    Consider an LPV-SA system $\Sigma\in\Sigma_{\dnp,\dnw}$ with behavior~$\mf{B}$. Given $(w_\mr{i},p_\mr{i})\in(\mb{R}^{\dnw}\times\mb{P})^{[1,T_\mr{i}]}$. If $T_\mr{i}\geq\LBf$ and
    \begin{equation}\label{eq:lem:state:concaternation}
        (w_\mr{i},p_\mr{i})\land(w_\mr{r}, p_\mr{r})\in\Bfint{}{[1,T_\mr{i}+T_\mr{r}]},
    \end{equation}
    then, the initial condition of $(w_\mr{r}, p_\mr{r})$ is uniquely expressed in terms of $(w_\mr{i},p_\mr{i})$. {Equivalently}, for a minimal LPV-SS representation of $\Sigma$, there exists a unique $\mt{x}\in\mb{R}^{\nBf}$ that serves as the initial condition of $(w_\mr{r}, p_\mr{r})$. 
\end{lemma}
\begin{proof}
    See Appendix~\ref{app:pf:lemstate}.
\end{proof}
 The result of Lemma~\ref{lem:state} provides us with a condition ($T_\mr{i}\geq\LBf$) that ensures a unique\footnote{The formal problem formulation in Section~\ref{ss:simprob} does \emph{not} require {finding} a unique response for $y_\mr{r}$, which is, however, often desired.} response to an input-scheduling pair. %
In the next section, we give a solution to the LPV data-driven simulation problem.

\subsection{The LPV data-driven simulation algorithm}

To satisfy $(w_\mr{i},p_\mr{i})\land(\col(u_\mr{r}, y_\mr{r}), p_\mr{r})\in\Bfint{}{[1,T_\mr{i}+T_\mr{r}]}$, we know from Theorem~\ref{thm:FLeasy} %
that~\eqref{eq:thm:dim} and thus~\eqref{eq:GPE} should hold for $L=T_\mr{i}+T_\mr{r}$. This allows to write
\[ \left[\begin{array}{c} \vphantom{\Big)} \mc{H}_{\bar{T}}(\breve{w}_{[1,\Nd]}) \\\hdashline \vphantom{\Big)}\mc{H}_{\bar{T}}(\breve{w}^{\breve{\mt{p}}}_{[1,\Nd]})-\mc{P}^\dnw_\mr{i,r}\mc{H}_{\bar{T}}(\breve{w}_{[1,\Nd]}) \end{array}\right] g = \left[\begin{array}{c}
            \mr{vec}(w_\mr{i}) \\ \mr{vec}(u_\mr{r}) \\ \mr{vec}(y_\mr{r}) \\\hdashline 0
        \end{array}\right], \]
where $\bar{T}=T_\mr{i}+T_\mr{r}$ and $\mc{P}^\dnw_\mr{i,r}= (p_\mr{i}\land p_\mr{r})\bkron I_\dnw$. For a given $w_\mr{i}$, $u_\mr{r}$, and scheduling sequence $p_\mr{i}\land p_\mr{r}$, this is a linear set of equations in the unknowns $g$ and $y_\mr{r}$. Partitioning of the Hankel matrices on the left-hand side provides Algorithm~\ref{alg:sim},  accomplishing LPV data-driven simulation. %
\begin{algorithm}[t]\caption{LPV data-driven simulation}\label{alg:sim}
    \begin{algorithmic}[1]
    \Require A data set $\dataset\in\Bfint{}{[1,\Nd]}$, an initial trajectory $(w_\mr{i},p_\mr{i})\in\Bfint{}{[1,T_\mr{i}]}$, and an input-scheduling trajectory $(u_\mr{r}, p_\mr{r})\in\pi_\mr{u,p}\Bfint{}{[1,T_\mr{r}]}$.
    \State Compute a $g$ that satisfies
    \begin{equation}\label{eq:alg:g}
        \begin{bmatrix} \mc{H}_{T_\mr{i}}(\breve{w}_{[1,\Nd-T_\mr{r}]}) \\ \mc{H}_{T_\mr{r}}(\breve{u}_{[T_\mr{i}+1,\Nd]}) \\ \mc{H}_{\bar{T}}(\breve{w}^{\breve{\mt{p}}}_{[1,\Nd]})-\mc{P}^\dnw_\mr{i,r}\mc{H}_{\bar{T}}(\breve{w}_{[1,\Nd]}) \end{bmatrix} g = \begin{bmatrix}
            \mr{vec}(w_\mr{i}) \\ \mr{vec}(u_\mr{r}) \\ 0
        \end{bmatrix},
    \end{equation}
    with $\bar{T}=T_\mr{i}+T_\mr{r}$ and $\mc{P}^\dnw_\mr{i,r}= (p_\mr{i}\land p_\mr{r})\bkron I_\dnw$. 
    \State Compute $y_\mr{r}$ via 
    \begin{equation}\label{eq:alg:y}
        \mr{vec}(y_\mr{r})= \mc{H}_{T_\mr{r}}(\breve{y}_{[T_\mr{i}+1,\Nd]})g.
    \end{equation}
    \Ensure $y_\mr{r}$
    \end{algorithmic}
\end{algorithm}
The LPV generalizations of the special cases of LTI data-driven simulation discussed in~\cite{MarkovskyRapisarda2008}, e.g., zero input response, zero initial condition response, impulse response, etc., directly follow from the %
{presented derivation of}
the LPV data-driven simulation algorithm. This {is} also %
the case for, e.g., recursive implementation of Algorithm~\ref{alg:sim} to simulate for $T_\mr{r}\to\infty$, data-driven simulation for LPV embeddings of nonlinear systems via iterative scheduling refinement,\footnote{\label{footnote:iterative}{In the case of nonlinear systems, the scheduling is generally dependent on $w$ through a so-called scheduling map. In~\cite{CiWe20}, an iterative procedure in the context of model predictive control is used to obtain the $w$-dependent scheduling signal, {similar to the approach used in}
sequential quadratic programming. See also Example~\ref{exmp:nl} for application of the iterative scheme in a nonlinear data-driven control setting.}} or data-based scheduling estimation. {While} these topics are interesting, {due to the sake of space, they are} %
not %
discussed {in detail in this paper}.

\section{Examples}\label{s:examples}
To illustrate the validity and effectiveness of the data-driven representations, we demonstrate their capabilities on an LPV system and an LPV embedding of a nonlinear system in a simulation and control scenario.

\begin{example}
For this first example, we use the MSD system presented in Example~\ref{exmp:msd} with %
parameters given in Table~\ref{tab:prm}.
\begin{table}
    \centering
    \caption{Parameters of the MSD system.}\label{tab:prm}
    \begin{tabular}{|c|c|c|c|c|}
        \hline $m$ & $\springconstant_0$ & $\springconstant_1$ & $d$ & $T_\mr{s}$ \\\hline $25$ [kg] & $5.5$ [N/m] & $4.5$ [N/m] & $1$ [Ns/m]  & $0.1$ [s]\\\hline
    \end{tabular}
\end{table}
Note that the system is SISO and $\LBf=\nBf=2$.
In this example, we want to perform a data-driven simulation of this system for $T_\mr{r}=35$ samples (corresponding to $3.5$ seconds), without having access to~\eqref{eq:msd} or its parameters; only a measured data set from the system is available. We consider three cases:
\begin{enumerate}[label={Case \arabic*:}, align=left, ref={\arabic*}, leftmargin=*]
    \item Conditions of Theorem~\ref{thm:FLeasy} hold and $T_\mr{i}\geq\LBf$;\label{case:good}
    \item \label{case:lessNd} Conditions of Theorem~\ref{thm:FLeasy} do not hold and $T_\mr{i}\geq\LBf$; 
    \item Conditions of Theorem~\ref{thm:FLeasy} hold and $T_\mr{i}<\LBf$. \label{case:lessTini}
\end{enumerate}
For the Cases~\ref{case:good} and~\ref{case:lessNd}, $T_\mr{i}=5$ {is chosen}, while, for Case~\ref{case:lessTini},  $T_\mr{i}=1$. We generate a data-dictionary~$\dataset$ of length $\Nd=(1+\dnw\dnp+\mBf)L+\nBf-1=161$, cf.~\eqref{eq:minimumNd}, and take, for Case~\ref{case:lessNd}, $\Nd=151$ by disregarding the last $10$ samples in $\dataset$. The data-dictionary used in this example is shown in Fig.~\ref{fig:ddict}, {where $\breve{u}(k)\sim\mc{N}(0,1)$ and $\breve{p}(k)\sim\mc{N}(0,1)$}. 
\begin{figure}
    \centering
    \includegraphics[width=0.975\linewidth]{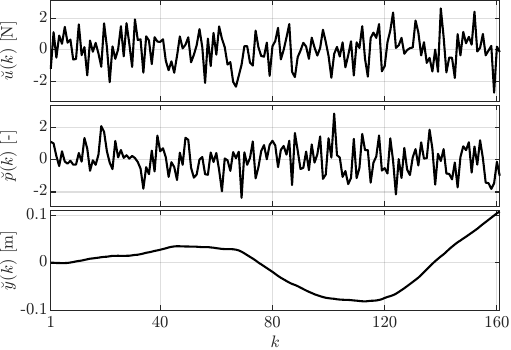}
    \caption{Data-dictionary measured from the LPV MSD system with $\Nd=161$.}
    \label{fig:ddict}
\end{figure}
With~\eqref{eq:GPE}, it is verified that~$\dataset$ satisfies the LPV-GPE condition and can represent the behavior on the horizon $T_\mr{i}+T_\mr{r}$ for Cases~\ref{case:good} and~\ref{case:lessTini}, while the LPV-GPE condition does not hold for Case~\ref{case:lessNd}.

We now solve Algorithm~\ref{alg:sim} for the aforementioned three cases. By solving~\eqref{eq:alg:g} in the least-squares sense and observe that, as expected, the following norm
\[ \left\|\begin{bmatrix} \mc{H}_{T_\mr{i}}(\breve{w}_{[1,\Nd-T_\mr{r}]}) \\ \mc{H}_{T_\mr{r}}(\breve{u}_{[T_\mr{i}+1,\Nd]}) \\ \mc{H}_{\bar{T}}(\breve{w}^{\breve{\mt{p}}}_{[1,\Nd]})-\mc{P}^\dnw_\mr{i,r}\mc{H}_{\bar{T}}(\breve{w}_{[1,\Nd]}) \end{bmatrix} g - \begin{bmatrix}
            \mr{vec}(w_\mr{i}) \\ \mr{vec}(u_\mr{r}) \\ 0
        \end{bmatrix}\right\|_2\]
is zero for Cases~\ref{case:good} and~\ref{case:lessTini}, while, for Case~\ref{case:lessNd}, {it} %
is 
$0.58$. Computing the simulated outputs with~\eqref{eq:alg:y} for {all the considered} cases  gives the results in Fig.~\ref{fig:cases}, 
\begin{figure}
    \centering
    \includegraphics[width=0.975\linewidth]{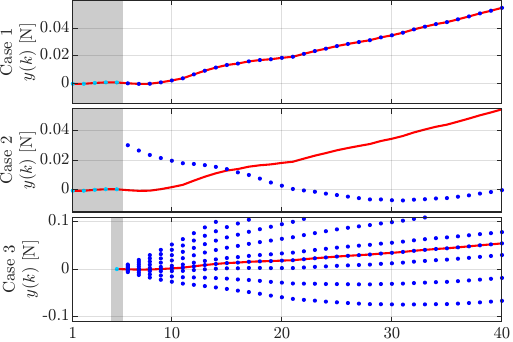}
    \caption{Simulation results for Cases~\ref{case:good}--\ref{case:lessTini}. The red solid line is the model-based simulation, the initial trajectory is indicated with the light-blue dotted line and the gray shaded area, while the blue dotted line represents the data-driven simulation result. For Case~\ref{case:good} the model-based and data-based simulations coincide. Case~\ref{case:lessNd} results in a simulated output that is not connected to the initial trajectory and hence does not coincide with the model-based simulation. The data-driven simulation results for Case~\ref{case:lessTini} show a continuation of the initial trajectory, but the simulated output trajectory is not unique which is illustrated by plotting several valid solutions for $y_\mr{r}$.}
    \label{fig:cases}
\end{figure}
which are compared to a model-based simulation of~\eqref{eq:msd} shown in red. In Case~\ref{case:good}, all the conditions for LPV data-driven simulation are satisfied, and indeed the model-based and data-based simulations coincide. The simulation result for Case~\ref{case:lessNd} shows that $y_\mr{r}$ is both not connected to the initial trajectory and not coinciding with the model-based simulation, i.e., the behavior, which contains the true trajectory, is not represented by the data in the smaller~$\dataset$. The solution to~\eqref{eq:alg:g} with Case~\ref{case:lessTini} results in  %
a larger space for
the valid $g$ {vectors} and thus a set of possible $y_\mr{r}$ {trajectories}. This is because with $T_\mr{i}<\LBf$, the problem~\eqref{eq:alg:g} is under-determined, i.e., there are infinitely many solutions for $y_\mr{r}$ for which $(\col(u_\mr{i}, y_\mr{i}), p_\mr{i})\land(\col(u_\mr{r}, y_\mr{r}), p_\mr{r})\in\Bfint{}{[1,T_\mr{i}+T_\mr{r}]}$. We have illustrated this by plotting a number of valid solutions for~$y_\mr{r}$, all of which are a valid continuation of the initial trajectory.
\end{example}
{
\begin{example}\label{exmp:nl}
In this next example, we apply our methods on a nonlinear system that is embedded {in} %
 an LPV {form.} 
 Consider the nonlinear system
\begin{multline}\label{eq:nlsys}
    y(k) + \big(0.2 - 0.4\tanh(y(k-1))\big)y(k-1) \\+ \tanh(y(k-2))y(k-2) = 1.2u(k-1) \\ + 0.4\sin(u(k-1))e^{-y^2(k-1)} \\ + \big(1+0.6\tanh(y(k-2))\big)y(k-2),
\end{multline}
and define $p:=\psi(u, y) = [ \, \tanh(y) \ \mr{sinc}(u)e^{-y^2} \, ]^\top$. This \emph{scheduling map} defines an LPV embedding of~\eqref{eq:nlsys} that has the form~\eqref{eq:lpvker-shiftedaff}. We will now use Algorithm~\ref{alg:sim} in a control setting (by seeing $w_\mr{r}$ as a decision variable) to achieve direct data-driven feedforward control of the nonlinear system~\eqref{eq:nlsys} under the assumption that we know $\psi$ (see \cite{verhoek2024kernelnote} on how to overcome this assumption). To achieve this, we will make use of Footnote~\ref{footnote:iterative}. We want to regulate the system~\eqref{eq:nlsys} to its origin in $T_\mr{r}=30$ time-steps from an arbitrary initial trajectory of length $T_\mr{i}=3$, {without knowing anything about the system, except a given data set recorded a priori.} %
{To obtain the data set $\dataset$,} we generate $\Nd=139$ {data} samples, cf.~\eqref{eq:minimumNd}, by applying an input $u(k)\sim\mc{N}(0,1)$ to~\eqref{eq:nlsys}. The resulting trajectories are shown in Fig.~\ref{fig:nlex_ddic}.
\begin{figure}
    \centering
    \includegraphics[width=0.975\linewidth]{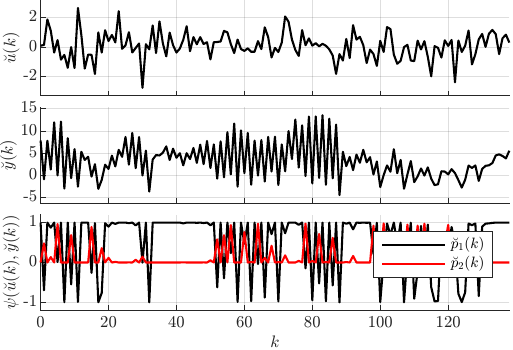}
    \caption{Input, output, and scheduling sequences {of} $\dataset$ {in} Example~\ref{exmp:nl}. The scheduling $p$ is defined by the scheduling map $\psi$.}
    \label{fig:nlex_ddic}
\end{figure}
A posteriori verification of~\eqref{eq:GPE} gives that $\dataset$ can represent $\Bfint{}{[1,T_\mr{i}+T_\mr{r}]}$. 

{Now, by exploiting} Algorithm~\ref{alg:sim}, {we formulate the solution of a predictive} %
control problem, {i.e., computing the required input sequence on a given horizon to drive the nonlinear system from arbitrary initial conditions to the origin.} Specifically, we iteratively solve the following quadratic program:
\begin{subequations}\label{eq:control}
\begin{align}
    \min_g \quad & (*)^\top Q \,\mr{vec}(y_\mr{r}) + (*)^\top R \,\mr{vec}(u_\mr{r}) \\
    \text{s.t.}\quad & \begin{bmatrix} \vphantom{\Big)} \mc{H}_{\bar{T}}(\breve{w}_{[1,\Nd]}) \\ \vphantom{\Big)}\mc{H}_{\bar{T}}(\breve{w}^{\breve{\mt{p}}}_{[1,\Nd]})-\mc{P}^\dnw_\mr{i,r}\mc{H}_{\bar{T}}(\breve{w}_{[1,\Nd]}) \end{bmatrix} g = \begin{bmatrix} \mr{vec}(w_\mr{i}) \\ \mr{vec}(u_\mr{r}) \\ \mr{vec}(y_\mr{r}) \\ 0
        \end{bmatrix},
\end{align}
\end{subequations}
where {in} each iteration, {after solving \eqref{eq:control}}, we update $\mc{P}^\dnw_\mr{i,r}$ using the scheduling map $\psi$ {applied on} the solution $(u_\mr{r}, y_\mr{r})$. 

{This procedure is summarized} in Algorithm~\ref{alg:con}.
\begin{algorithm}[t]\caption{Data-driven control for nonlinear systems}\label{alg:con}
    \begin{algorithmic}[1]
    \Require A data set $\dataset\in\Bfint{}{[1,\Nd]}$ for which~\eqref{eq:GPE} holds, an initial trajectory $(w_\mr{i},p_\mr{i})\in\Bfint{}{[1,T_\mr{i}]}$, and an initial guess for $p_\mr{r}$.
    \State\textbf{Repeat}
    \State\qquad Let $\mc{P}^\dnw_\mr{i,r}\gets(p_\mr{i}\land p_\mr{r})\bkron I_\dnw$,
    \State\qquad Solve~\eqref{eq:control} and obtain $(u_\mr{r},y_\mr{r})$
    \State\qquad Update $p_\mr{r}\gets\psi(u_\mr{r},y_\mr{r})$
    \State\textbf{Until} $p_\mr{r}$ has converged
    \Ensure $u_\mr{r}$
    \end{algorithmic}
\end{algorithm}
In~\cite{hespe2021convergence} it is {shown that with some additional minor modifications, this} %
sequence of quadratic programs has a local contraction property under assumptions similar to those for standard sequential quadratic programming methods. 

In this example, we choose $Q=R=I$ and take a zero-trajectory as initial guess for $p_\mr{r}$. We solve Algorithm~\ref{alg:con} for the initial trajectory
\begin{align*}
    w_\mr{i} &= \left(\begin{bsmallmatrix} 0.75 \\ 2.84 \end{bsmallmatrix},\begin{bsmallmatrix} -0.26 \\ 7.31 \end{bsmallmatrix},\begin{bsmallmatrix} -0.03 \\ 2.25 \end{bsmallmatrix}\right), \\ p_\mr{i} &= \left(\begin{bsmallmatrix} 0.99 \\ 0.00 \end{bsmallmatrix},\begin{bsmallmatrix} 0.99 \\ 0.00 \end{bsmallmatrix},\begin{bsmallmatrix} 0.978 \\ 0.006 \end{bsmallmatrix}\right),
\end{align*}
and took $\|p_\mr{r}^{n-1}-p_\mr{r}^{n}\|<10^{-6}$ as the stopping criterion for {Step}~5 of Algorithm~\ref{alg:con} with $p_\mr{r}^{n}$ the updated scheduling sequence after the $n$\tss{th} iteration. The trajectory for $p_\mr{r}$ converged in 12 steps. The difference between $y_\mr{r}$ coming from~\eqref{eq:control} and the true simulation with the implemented optimal $u_\mr{r}$ (in the 2-norm) is $1.5\cdot10^{-7}$. We have plotted the results in Fig.~\ref{fig:nlex_con}. 
\begin{figure}
    \centering
    \includegraphics[width=0.975\linewidth]{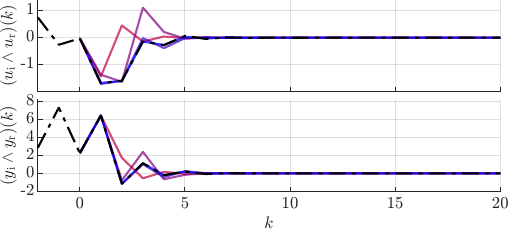}
    \caption{Input and output sequences for different iterations in Algorithm~\ref{alg:con} {depicted with} blue (less transparent lines correspond to a higher iteration number). The final input trajectory obtained %
   from Algorithm~\ref{alg:con} and the resulting output trajectory from the true system {are} depicted in black.}
    \label{fig:nlex_con}
\end{figure}
This figure also shows  $(u^n_\mr{r},y^n_\mr{r})$, i.e., the optimized trajectories for iteration~$n$ in {Step}~3 of Algorithm~\ref{alg:con}. The iterations are plotted in the color spectrum from red (small~$n$) and blue (large~$n$). These results show that the simulated trajectories of converge to the true solution of the nonlinear system relatively fast, by simply solving a number of simple quadratic programs. {This highlights how the derived results can be applied to establish} %
efficient direct data-driven analysis and control methods for nonlinear systems. 
\end{example}
}

\section{Conclusions}\label{s:con}

In this paper, we derived {foundational results for data-driven LPV analysis and control of LPV systems in terms of establishing a computable} data-driven representation {of LPV systems together with a generalized form of a persistency of excitation condition. The latter condition characterizes when the data is sufficiently informative to describe the underlying behavior of the system via such a data-driven representation.} Additionally, we provided a formal solution to the LPV data-driven simulation problem {to compute responses of LPV systems for a given input and scheduling trajectory when no model based knowledge is available for the system except previously recorded data. We demonstrated validity of our results through various examples, also showing applicability of the proposed approaches for data-driven handling of nonlinear systems.}

{In our work, we restricted the scope to} 
 LPV-SA systems, i.e., systems that can be represented by an LPV-IO representation with shifted-affine scheduling dependence, {which is a practically useful class of systems. However, it is an important objective of future research} 
to generalize the results to other useful scheduling dependency classes and to handling noisy data sets. Moreover, we consider the formulation of conditions on the input and scheduling trajectories that allow for experiment design with a priori guarantees on the satisfaction of~\eqref{eq:GPE} an important topic for future research.

\bibliographystyle{IEEEtran}
\bibliography{refs_flpaper}
\appendices
\section{Proof of Lemma~\ref{lem:dim}}\label{app:pf:lemdim}
Before we can give the proof for Lemma~\ref{lem:dim}, we need some intermediate results. We first write the LPV-SS representation~\eqref{eq:lpvssgen} with~\eqref{eq:lpvss} in a structured form. This, in turn, allows us to draw conclusions on {the} observability of~\eqref{eq:lpvssgen} with~\eqref{eq:lpvss}. Finally, these are used to give the proof of Lemma~\ref{lem:dim}.
\subsection{Structured LPV-SS form}\label{aa:structuredss}
Through the state construction~\eqref{eq:stateconstruction} and the affine parametrization of the scheduling dependent polynomial coefficients of the kernel/IO representations, we can write the LPV-SS representation in terms of $x(k)$, $u(k)$, $p(k)\kron y(k)$, $p(k)\kron u(k)$, and $p(k)\kron p(k)\kron u(k)$ by splitting up~\eqref{eq:lpvss} {into} scheduling-independent and scheduling-dependent parts:
\begin{subequations}\label{eq:SSrep_shiftaff}
\begin{align}
    \hspace{-2.0mm}x({k+1}) & = A_0 x(k) +  B_0 u(k) + A_\mr{p} p(k)\kron x_1(k) \notag\\ 
    & \hphantom{=\;}  + B_\mr{p} p(k)\kron u(k) + B_\mr{pp} p(k)\kron p(k)\kron u(k), \label{eq:SSrep_shiftaff:state} \hspace{-2mm}\\
y(k) & = C x(k) + D_0 u(k) + D_\mr{p} p(k)\kron u(k), \label{eq:SSrep_shiftaff:output}
\end{align}
\end{subequations}
with $A_0, \dots, D_\mr{p}$ as in~\eqref{eq:matrices_structured} on the next page.
\begin{figure*}
\begin{subequations}\label{eq:matrices_structured}
\begin{align}
		A_0 & = \begin{bsmallmatrix}
		-a_{1,0} & I  & \hdots & 0 \\
		\svdots & \svdots & \sddots & \svdots \\
		-a_{n_\mr{a}-1,0} & 0 & \hdots & I \\
		-a_{n_\mr{a},0} & 0 & \hdots & 0 
		\end{bsmallmatrix}, &
		B_0 & = \begin{bsmallmatrix}
		b_{1,0} - a_{1,0}b_{0,0} \\
		\svdots \\
		b_{n_\mr{b}-1,0} - a_{n_\mr{b}-1,0}b_{0,0} \\
		b_{n_\mr{b},0} - a_{n_\mr{b},0}b_{0,0}
		\end{bsmallmatrix}, &
		B_\mr{pp} & = \begin{bsmallmatrix}
		-a_{1,1}b_{0,1} & \hdots & -a_{1,n_\mr{p}}b_{0,n_\mr{p}}\\
		\svdots & \sddots & \svdots \\
		-a_{n_\mr{a},1}b_{0,1} & \hdots & -a_{n_\mr{a},n_\mr{p}}b_{0,n_\mr{p}}
		\end{bsmallmatrix}, &
		D_0 & = b_{0,0}, \\
		A_\mr{p} & = \begin{bsmallmatrix}
		-a_{1,1} & \hdots & -a_{1,n_\mr{p}}\\
		\svdots & \sddots & \svdots \\
		-a_{n_\mr{a},1} & \hdots & -a_{n_\mr{a},n_\mr{p}}
		\end{bsmallmatrix}, &
		B_\mr{p} & = \begin{bsmallmatrix}
		b_{1,1} & \hdots & b_{1,n_\mr{p}}\\
		\svdots & \sddots & \svdots \\
		b_{n_\mr{b},1} & \hdots & b_{n_\mr{b},n_\mr{p}}
		\end{bsmallmatrix}, &
		C & = \begin{bmatrix}
		I & 0 & \cdots & 0
		\end{bmatrix}, &
		D_\mr{p} & = \begin{bsmallmatrix}
		    b_{0,1} & \hdots & b_{0,n_\mr{p}}
		\end{bsmallmatrix}
\end{align}
\end{subequations}
\hrule
\end{figure*}
Note that $x_1(k)$ is as in~\eqref{eq:stateconstruction:x1}, i.e., $x_1(k)\in\mb{R}^\dny$. Formulation~\eqref{eq:SSrep_shiftaff} separates the LTI behavior from the parameter-varying behavior. {By substituting $x_1$ in $p(k)\kron x_1(k)$ in~\eqref{eq:SSrep_shiftaff:state} with \eqref{eq:stateconstruction:x1}, the
state equation is written as} %
\begin{multline}
    x(k+1) = A_0x(k) + B_0 u(k) + A_\mr{p} p(k)\kron y(k)  \\ +\tilde{B}_\mr{p} p(k)\kron u(k) + \tilde{B}_\mr{pp}  p(k)\kron p(k)\kron u(k),\label{eq:stateeq_intermsof_y}
\end{multline}
where $\tilde{B}_\mr{p}=B_\mr{p} - A_\mr{p}(I_\dnp\kron D_0)$ and $\tilde{B}_\mr{pp}=B_\mr{pp} - A_\mr{p}(I_\dnp\kron D_\mr{p})$.
With this form, we can express a length~$L$ trajectory in $\Bfint{}{[1,L]}$ as follows
\begin{subequations}\label{eq:writteninoutputform}
\begin{multline}\label{eq:hugeoutputequation2}
    \mr{vec}(y_{[1,L]}) = \ms{O}_{L}\,\mt{x} + \ms{T}_L\mr{vec}(u_{[1,L]})+\ms{O}^{\mt{p}}_L\mr{vec}(y^{\mt{p}}_{[1,L]}) \\+\ms{T}^{\mt{p}}_L\mr{vec}(u^{\mt{p}}_{[1,L]}) + \ms{T}^{\mt{pp}}_L\mr{vec}(u^{\mt{pp}}_{[1,L]}),
\end{multline}
with $u_{[1,L]}^\mt{pp}=\big(p(k)\kron p(k)\kron u(k)\big)_{k=1}^L$ and  initial condition $\mt{x}\in\mb{R}^{\dnx}$, where
\begin{align}
    \ms{O}_L & =  \begin{bsmallmatrix}C \\ CA_0 \\ \svdots\\ CA_0^{L-1} \end{bsmallmatrix}, \qquad 
    \ms{T}_L = \begin{bsmallmatrix}
		D_0 & 0 & \cdots  & 0 \\
		CB_0 & D_0 & \sddots   & \svdots \\
		\svdots & \sddots & \sddots   & 0 \\
		CA_0^{L-1}B_0  & \cdots & CB_0 & D_0 
		\end{bsmallmatrix}, \notag\\ 
    \ms{O}^{\mt{p}}_L & = \begin{bsmallmatrix}
		0 & 0 & \cdots  & 0 \\
		CA_\mr{p} & \sddots & \sddots  & \svdots \\
		\svdots & \sddots & \sddots  & 0 \\
		CA_0^{L-1}A_\mr{p} & \cdots & CA_\mr{p} & 0 
		\end{bsmallmatrix},  \notag\\ 
	\ms{T}^{\mt{p}}_L & =\begin{bsmallmatrix}
		D_\mr{p} & 0 & \cdots  & 0 \\
		C\tilde{B}_\mr{p} & D_\mr{p} & \sddots  & \svdots \\
		\svdots & \sddots & \sddots  & 0 \\
		CA_0^{L-1}\tilde{B}_\mr{p}  & \cdots & C\tilde{B}_\mr{p} & D_\mr{p}
		\end{bsmallmatrix}, \notag\\
    \ms{T}^{\mt{pp}}_L&=\begin{bsmallmatrix}
		0 & 0 & \cdots  & 0 \\
		C\tilde{B}_\mr{pp} & \sddots & \sddots & \svdots \\
		\svdots & \sddots & \sddots & 0 \\
		CA_0^{L-1}\tilde{B}_\mr{pp} & \cdots & C\tilde{B}_\mr{pp} & 0
		\end{bsmallmatrix}.\label{eq:matrices}
\end{align}
\end{subequations}
With these formulations, we now provide observability and minimality properties of the LPV-SS representation, which are required for the proof of Lemma~\ref{lem:dim}.

\subsection{Observability %
of the structured LPV-SS {form}}
We first introduce the notion of \emph{complete state-observability} from~\cite{Toth2010}:
\begin{definition}[Complete state-observability~\cite{Toth2010}]
    An LPV-SS representation~\eqref{eq:lpvssgen} is called \emph{completely state-observable}, if for all $(u,x,y,p)\in\Bss$, $(u,x^\prime,y,p)\in\Bss$ it holds that $x=x^\prime$.
\end{definition}
From which, the next result follows:
\begin{lemma}\label{lem:complete-obs}
    {Given a shifted-affine LPV-IO realization~\eqref{eq:lpvio-shifted} for which the polynomials in~\eqref{eq:abpoly} are left-coprime. The LPV-SS representation {with matrices}~\eqref{eq:lpvss} that is constructed from this LPV-IO representation} is completely state-observable. 
\end{lemma}
\begin{proof} 
    Consider~\eqref{eq:writteninoutputform} {for some} $L\geq\dnx$. Computing $\ms{O}_L$ for this $L$ gives
    \[ \ms{O}_L = \begin{bmatrix}
        I & 0 & \cdots  & \cdots & \cdots & 0 \\ -a_{1,0} & I & 0 & \cdots &\cdots & 0 \\ a^2_{1,0}-a_{2,0} & -a_{1,0} & I & 0 & \cdots & 0 \\ 
        \vdots & \vdots & \vdots & \vdots & \vdots & \vdots \\
    \end{bmatrix}, \] 
    i.e., $\ms{O}_L$ is a tall, lower-triangular matrix with~1's on the diagonal, i.e., $\ms{O}_L$ is full column-rank. Hence, for a given $(u_{[1,L]},y_{[1,L]},p_{[1,L]})\in\pi_\mr{u,y,p}\Bss$, {the initial state} $\mt{x}$ {can be uniquely determined {via}~\eqref{eq:writteninoutputform}, after which the state trajectory is governed by~\eqref{eq:stateeq_intermsof_y}. This implies that the state trajectory for $k\geq 1$ is unique. {Since this holds for every $L\geq\dnx$}, together with the time-invariance property of LPV systems, the state trajectory $x$  {corresponding  to each} $(u,y,p)\in\pi_\mr{u,y,p}\Bss$ is unique, i.e.,} the representation is  completely state-observable.
\end{proof}
This result states that for any scheduling sequence in~$\mf{B}_\mb{P}$, the representation~\eqref{eq:lpvss} is observable. We have already established in Section~\ref{sss:lpvsses} that we can always obtain a minimal LPV-SS {representation} with static scheduling dependence {for the considered behaviors}. Note that this minimal realization will admit the same structure as discussed in Section~\ref{aa:structuredss}, i.e., we can always formulate~\eqref{eq:writteninoutputform} with an~$\mt{x}$ of dimension~$\nBf$, where the matrices $\ms{O}_{T_\mr{r}}, \dots, \ms{T}^{\mt{pp}}_{T_\mr{r}}$ are constructed as in~\eqref{eq:matrices}.

Finally, we have the following result, which is key in deriving the proof for Lemma~\ref{lem:dim}. This result links the rank of the `LTI part' of the observability matrix, i.e., $\ms{O}_L$ to the invariant integer $\LBf$.
\begin{proposition}\label{prop:OL-FR-iff-L>L(B)}
    Given a minimal~\eqref{eq:lpvss} constructed from~\eqref{eq:lpvio-shifted} where {the polynomials in}~\eqref{eq:abpoly} are left-coprime. Then $\mr{rank}(\ms{O}_L)=\nBf$, if and only if $L\geq\LBf$.
\end{proposition}
\begin{proof}
    First note that by Lemma~\ref{lem:complete-obs}, the representation is observable for \emph{any} $p\in\mf{B}_\mb{P}$, including $0\in\mf{B}_\mb{P}$. Then it follows from \cite[Thm.~6]{willems1986time} and \cite[Sec.~4.3]{Toth2010} that the observability index and $\dnr$, i.e., $\LBf$ are equal. This implies that $\mr{rank}(\ms{O}_{L})= \nBf$, if and only if $L\geq\dnr = \LBf$, concluding the proof.
\end{proof}
We are now ready to give the proof of Lemma~\ref{lem:dim}.

\subsection{Proof of Lemma~\ref{lem:dim}}

\begin{proof} Take an arbitrary ${p}_{[1,L]}\in\Bfint{\mb{P}}{[1,L]}$ and an associated $\col({u}_{[1,L]}, y_{[1,L]}) \in\Bfint{p}{[1,L]}$. Now consider the formulation of this trajectory in terms of~\eqref{eq:writteninoutputform}.
We can write
\[ \mr{vec}(u^{\mt{p}}_{[1,L]}) = (p_{[1,L]}\bkron I_{\dnu})\mr{vec}(u_{[1,L]}),\]
similarly for $\mr{vec}(y_{[1,L]}^{\mt{p}})$. Furthermore, note that $\mr{vec}(u^{\mt{pp}}_{[1,L]})$ can be written as 
\[\mr{vec}(u^{\mt{pp}}_{[1,L]})=(p_{[1,L]}\bkron I_{\dnu\dnp})\mr{vec}(u^\mt{p}_{[1,L]}).\]
Hence, with $\mc{P}^{\dnu}=p_{[1,L]}\bkron I_{\dnu}$, $\mc{P}^{\dny}=p_{[1,L]}\bkron I_{\dny}$, and $\mc{P}^{\dnu\dnp}=p_{[1,L]}\bkron I_{\dnu\dnp}$, we can rewrite~\eqref{eq:writteninoutputform} with $\mt{x}\in\mb{R}^{\nBf}$ as
\begin{multline}\label{eq:LPVIOequationform_ss}
	\left(I-\ms{O}^{\mt{p}}_L\mc{P}^{\dny}\right)\mr{vec}(y_{[1,L]}) = \ms{O}_{L}\mt{x} \\
	+ \left(\ms{T}_L + \ms{T}^{\mt{p}}_L\mc{P}^{\dnu} + \ms{T}^{\mt{pp}}_L\mc{P}^{\dnu\dnp}\mc{P}^{\dnu}\right) \mr{vec}(u_{[1,L]}).
\end{multline}
Hence, to characterize the manifest behavior of the LPV system for a given scheduling sequence $p_{[1,L]}$, we can use~\eqref{eq:LPVIOequationform_ss} to express the dynamic relation of any trajectory in $\Bfint{p}{[1,L]}$ with
\begin{multline}\label{eq:notkernelform2}
	\begin{bmatrix}
		I & 0 \\ 
		0 & I-\ms{O}^{\mt{p}}_L\mc{P}^{\dny} 
	\end{bmatrix}
	\mr{vec}(w_{[1,L]}) = \\
	\begin{bmatrix}
		0 & I  \\
		\ms{O}_L & \ms{T}_L + \ms{T}^{\mt{p}}_L\mc{P}^{\dnu} + \ms{T}^{\mt{pp}}_L\mc{P}^{\dnu\dnp}\mc{P}^{\dnu} 
	\end{bmatrix}
	\begin{bmatrix}
		\mt{x}  \\ \mr{vec}(u_{[1,L]})
	\end{bmatrix}.
\end{multline}
As $\ms{O}^{\mt{p}}_L$ is a strictly lower block-triangular matrix and $\mc{P}^{\dny}$ is a block diagonal matrix, $\ms{O}^{\mt{p}}_L\mc{P}^{\dny}$ is always strictly lower triangular. Therefore,
\begin{equation}\label{eq:pfdim:lhs}
    \begin{bmatrix}
		I & 0 \\ 
		0 & I-\ms{O}^{\mt{p}}_L\mc{P}^{\dny} 
	\end{bmatrix}
\end{equation}
is nonsingular. Thus, we can write
\begin{equation}
    \mr{vec}(w_{[1,L]}) = \underbrace{\begin{bmatrix}I & 0 \\ 0 & I-\ms{O}^{\mt{p}}_L\mc{P}^{\dny} \end{bmatrix}^{-1}\begin{bmatrix} 0 & I  \\ \ms{O}_L & \ms{Q} \end{bmatrix}}_{\mc{B}_p} \begin{bmatrix} \mt{x}  \\ \mr{vec}(u_{[1,L]}) \end{bmatrix},
\end{equation}
with $\ms{Q} = \ms{T}_L + \ms{T}^{\mt{p}}_L\mc{P}^{\dnu} + \ms{T}^{\mt{pp}}_L\mc{P}^{\dnu\dnp}\mc{P}^{\dnu}$. Since $w,p$ were chosen arbitrarily, we conclude that the columns of $\mc{B}_p$ form a basis for $\Bfint{p}{[1,L]}$. The dimension of this basis is purely governed by $\begin{bsmallmatrix} 0 & I \\ \ms{O}_L&\ms{Q}\end{bsmallmatrix}$, because \eqref{eq:pfdim:lhs} is nonsingular. To conclude the proof, observe that the LPV-SS representation is minimal and $L\geq\LBf$. {Hence,} Proposition~\ref{prop:OL-FR-iff-L>L(B)} gives that
\[ \rank \left(\begin{bmatrix} 0 & I \\ \ms{O}_L  &  \ms{Q} \end{bmatrix}\right)=\nBf+\mBf L, \]
i.e., $\rank(\mc{B}_p)=\nBf+\mBf L=\dim(\mr{image}(\mc{B}_p)) = \dim(\Bfint{p}{[1,L]})$.
\end{proof}

\section{Proof of Theorem~\ref{thm:FLeasy}}\label{app:pf:fl}
\begin{proof}
We first prove Item~\ref{thmitem:beh}~$\Leftrightarrow$~Item~\ref{thmitem:dim}.
First, note that $\Bfint{p}{[1,L]}$ is a linear subspace. From Lemma~\ref{lem:dim}, we know that for $L\geq\LBf$, $\mr{dim}(\Bfint{p}{[1,L]})= \nBf + \mBf L$. Hence, if~\eqref{eq:thm:beh} holds, then $\mr{dim}(\Bfint{p}{[1,L]})=\mr{dim}(\mr{image}(\mc{H}_L(\breve{w}_{[1,\Nd]})\meu{N}_{p}))= \nBf + \mBf L$, i.e.,~\eqref{eq:thm:dim} holds, which concludes the `\eqref{eq:thm:beh}$\Rightarrow$\eqref{eq:thm:dim}' direction.

Now we show the `\eqref{eq:thm:beh}$\Leftarrow$\eqref{eq:thm:dim}' direction. Consider~\eqref{eq:fundamentallemmafull} and an arbitrary scheduling sequence $\hat{p}_{[1,L]}\in\Bfint{\mb{P}}{[1,L]}$. For any $g$ that is both in the row space of $\mc{H}_L(\breve{w}_{[1,\Nd]})$ and in the kernel of $\mc{H}_L(\breve{w}^{\breve{\mt{p}}}_{[1,\Nd]}) - \hat{\mc{P}}^{\dnu}\mc{H}_L(\breve{w}_{[1,\Nd]})$, we obtain a trajectory $\hat{w}_{[1,L]}$. Based on the derivations in Section~\ref{sec:shiftedaff_ddrep}, we conclude that $(\hat{w}_{[1,L]},\hat{p}_{[1,L]})$ trivially satisfies~\eqref{eq:lpvker-shiftedaff}, i.e., $(\hat{w}_{[1,L]},\hat{p}_{[1,L]})\in\Bfint{}{[1,L]}$ and thus $\hat{w}_{[1,L]}\in\Bfint{\hat{p}}{[1,L]}$. Hence, for this fixed scheduling sequence $\hat{p}_{[1,L]}$, the space spanned by $\mr{image}(\mc{H}_L(\breve{w}_{[1,\Nd]})\meu{N}_{\hat{p}})$ is a subspace of $\Bfint{p}{[1,L]}$, i.e.,
\begin{equation}\label{eq:thm:pf:incl}
    \mr{image}(\mc{H}_L(\breve{w}_{[1,\Nd]})\meu{N}_{\hat{p}}) \subseteq \Bfint{p}{[1,L]}.
\end{equation}
Therefore, as the dimension of $\Bfint{p}{[1,L]}$ is equal to $\nBf+\mBf L$,  if~\eqref{eq:thm:dim} holds, {then}~\eqref{eq:thm:pf:incl} must hold with equality, i.e.,~\eqref{eq:thm:beh} must hold. The proof of Item~\ref{thmitem:rep} follows directly from the above reasoning and the derivations in Section~\ref{sec:shiftedaff_ddrep}. To prove Item~\ref{thmitem:gpe}, note that through~\eqref{eq:mBfprime}, the right-hand side of condition~\eqref{eq:GPE} is equivalent to the right-hand side of~\eqref{eq:GPE_ltiembedding}. Then, the proof follows from the fact that \eqref{eq:GPE_ltiembedding}$\Leftrightarrow$\eqref{eq:FLltiemb}, from which~\eqref{eq:fundamentallemmafull} is obtained without any loss of equivalence. Since Item~\ref{thmitem:rep}~$\Leftrightarrow$~Item~\ref{thmitem:dim}, we have that~\eqref{eq:fundamentallemmafull}$\Leftrightarrow$\eqref{eq:thm:dim}, which concludes the proof.
\end{proof}

\section{Proof of Lemma~\ref{lem:state}}\label{app:pf:lemstate}
\begin{proof}
Consider the direct LPV-SS realization of $\Sigma$ with state dimension $\nBf$ and behavior $\Bss$. Consider a trajectory $(u_{[k_1,k_2]}, y_{[k_1,k_2]},p_{[k_1,k_2]},x_{[k_1,k_2]})\in\left.\Bss\right|_{[k_1,k_2]}$, with $k_1\geq k_2$, and let us denote $\bar{k} = k_2-k_1$ and $y_{\bar{k}}:=\mr{vec}(y_{[k_1,k_2]})$ for brevity (similarly for $u$ and $p$).
From the discussion in Section~\ref{aa:structuredss}, we know that we can express $(u_{[k_1,k_2]}, y_{[k_1,k_2]},p_{[k_1,k_2]})\in\pi_\mr{u,y,p}\left.\Bss\right|_{[k_1,k_2]}$ in the form of~\eqref{eq:writteninoutputform}:
\begin{equation}
    (I - \ms{O}^{\mt{p}}_{\bar{k}}\mc{P}_{[k_1,k_2]}^\dny)y_{\bar{k}} = \ms{O}_{\bar{k}}x(k_1) + \ms{Q}_{\bar{k}}u_{\bar{k}}
\end{equation}
where 
\[ \ms{Q}_{\bar{k}} = \ms{T}_{\bar{k}} + \ms{T}^{\mt{p}}_{\bar{k}}\mc{P}_{[k_1,k_2]}^\dnu + \ms{T}^{\mt{pp}}_{\bar{k}} \mc{P}_{[k_1,k_2]}^{\dnu\dnp}\mc{P}_{[k_1,k_2]}^{\dnu}, \]
with $\mc{P}_{[k_1,k_2]}^{\dnu}=p_{[k_1,k_2]}\bkron I_{\dnu}$, $\mc{P}_{[k_1,k_2]}^{\dny}=p_{[k_1,k_2]}\bkron I_{\dny}$, and $\mc{P}_{[k_1,k_2]}^{\dnu\dnp}=p_{[k_1,k_2]}\bkron I_{\dnu\dnp}$. 
Furthermore, we can also express $x(k_2)$ by recursive application of~\eqref{eq:stateeq_intermsof_y} {starting} from $x(k_1)$:
\begin{multline}\label{eq:pfinit:k2}
    x(k_2) = A_0^{k_2-k_1} x(k_1) + \ms{A}_{\bar{k}}\mc{P}_{[k_1,k_2]}^\dny y_{\bar{k}} + \ms{B}_{\bar{k}} u_{\bar{k}} + \\ \ms{B}^\mt{p}_{\bar{k}}\mc{P}_{[k_1,k_2]}^{\dnu} u_{\bar{k}}
    + \ms{B}^\mt{pp}_{\bar{k}} \mc{P}_{[k_1,k_2]}^{\dnu\dnp}\mc{P}_{[k_1,k_2]}^{\dnu}u_{\bar{k}},
\end{multline}
where
\begin{align*}
    \ms{A}_{\bar{k}} & = \begin{bsmallmatrix} A_0^{\bar{k}-1}A_\mr{p} & \cdots & A_\mr{p} & 0 \end{bsmallmatrix}, &
    \ms{B}_{\bar{k}} & = \begin{bsmallmatrix} A_0^{\bar{k}-1}B_0 & \cdots & B_0 & 0 \end{bsmallmatrix}, \\
    \ms{B}^\mt{p}_{\bar{k}} & = \begin{bsmallmatrix} A_0^{\bar{k}-1}\tilde{B}_\mr{p} & \cdots & \tilde{B}_\mr{p} & 0 \end{bsmallmatrix}, &
    \ms{B}^\mt{pp}_{\bar{k}} & = \begin{bsmallmatrix} A_0^{\bar{k}-1}\tilde{B}_\mr{pp} & \cdots & \tilde{B}_\mr{pp} & 0 \end{bsmallmatrix}.
\end{align*}
We will now express $x(k_2)$ in terms of only the trajectories $(\col(u_{[k_1,k_2]}, y_{[k_1,k_2]}),p_{[k_1,k_2]})$. 
Suppose $\ms{O}_{\bar{k}}$ has a left-inverse~$\ms{O}_{\bar{k}}^{+}$, then we obtain the following expression for $x(k_1)$:
\begin{equation}\label{eq:pfinit:k1}
    x(k_1) = \ms{O}_{\bar{k}}^{+}(I - \ms{O}^{\mt{p}}_{\bar{k}}\mc{P}_{[k_1,k_2]}^\dny)y_{\bar{k}} -  \ms{O}_{\bar{k}}^{+}\ms{Q}_{\bar{k}}u_{\bar{k}}.
\end{equation}
Substitution of~\eqref{eq:pfinit:k1} in~\eqref{eq:pfinit:k2} gives
\begin{multline}\label{eq:pfinit:k_2full}
    x(k_2) = \left(A_0^{k_2-k_1} \ms{O}_{\bar{k}}^{+}(I - \ms{O}^{\mt{p}}_{\bar{k}}\mc{P}_{[k_1,k_2]}^\dny) + \ms{A}_{\bar{k}}\mc{P}_{[k_1,k_2]}^\dny \right) y_{\bar{k}} \\
    +  \Big(A_0^{k_2-k_1}\ms{O}_{\bar{k}}^{+}\ms{Q}_{\bar{k}} + \ms{B}_{\bar{k}} + \ms{B}^\mt{p}_{\bar{k}}\mc{P}_{[k_1,k_2]}^{\dnu} \cdots \\
    + \ms{B}^\mt{p}_{\bar{k}} \mc{P}_{[k_1,k_2]}^{\dnu\dnp}\mc{P}_{[k_1,k_2]}^{\dnu}\Big)u_{\bar{k}},
\end{multline}
i.e., $x(k_2)$ can be uniquely expressed from the trajectory $(\col(u_{[k_1,k_2]}, y_{[k_1,k_2]}),p_{[k_1,k_2]})$ given that $\ms{O}_{\bar{k}}$ has a left-inverse. 
From Proposition~\ref{prop:OL-FR-iff-L>L(B)}, we have that $\ms{O}_{\bar{k}}$ is full column rank, i.e., $\ms{O}_{\bar{k}}$ has a left-inverse, if and only if $\bar{k}\geq\LBf$. For $k_1=1$ and $k_2=T_\mr{i}+1$, $x(k_2)$ is the initial condition for the trajectory $(w_\mr{r}, p_\mr{r})$, which can be uniquely expressed in terms of $(w_\mr{i},p_\mr{i})$ and\footnote{Note that if there is no feed-through, i.e., $D(p)=b_0(p)=0$, then~\eqref{eq:pfinit:k_2full} can be expressed in terms of only $(w_{[k_1,k_2-1]}, p_{[k_1,k_2-1]})$.} $w(T_\mr{i}+1), p(T_\mr{i}+1)$ if and only if $\bar{k}=T_\mr{i}\geq\LBf$, concluding the proof.
\end{proof}

\ifx\tacversion\undefined
	\vspace*{1em}
	\renewcommand{\baselinestretch}{0.83}
	\selectfont
\fi

\makebiography{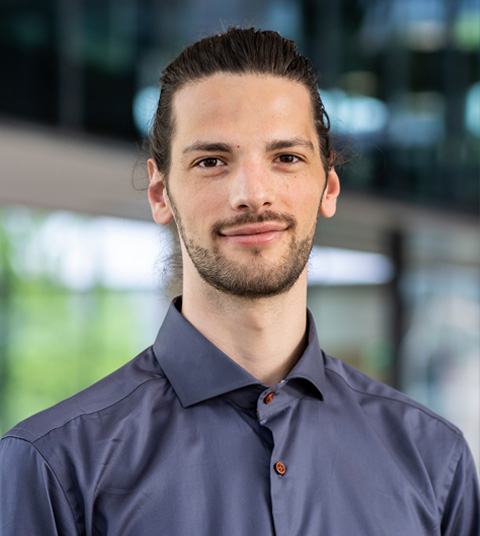}{Chris Verhoek}{received his M.Sc. degree (Cum Laude) in Systems and Control from the Eindhoven University of Technology (TU/e) in 2020. In 2025, he received his Ph.D. degree with Cum Laude distinction, also from the TU/e. His M.Sc. thesis was selected as best thesis of the Electrical Engineering department in the year 2020. 
He is currently a postdoctoral researcher with the Control Systems group, Dept. of Electrical Engineering, TU/e. In the fall of 2023, he was a visiting researcher at the IfA, ETH Z{\"u}rich, Switzerland. His main research interests include (data-driven) analysis and control of nonlinear and LPV systems and learning-for-control techniques with stability and performance guarantees.}

\makebiography{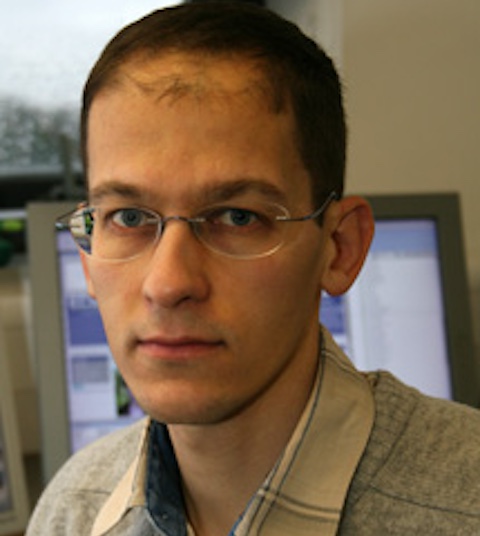}{Ivan Markovsky}{received the Ph.D. degree in electrical engineering from the Katholieke Universiteit Leuven, Leuven, Belgium, in February 2005. He is currently an ICREA Professor with the International Centre for Numerical Methods in Engineering, Barcelona. From 2006 to 2012, he was an Assistant Professor with the School of Electronics and Computer Science, University of Southampton, Southampton, U.K., and from 2012 to 2022, an Associate Professor with the Vrije Universiteit, Brussel, Belgium. His research interests are computational methods for system theory, identification, and data-driven control in the behavioral setting. Dr. Markovsky was the recipient of an ERC starting grant ``Structured low-rank approximation: Theory, algorithms, and applications'' 2010-2015, Householder Prize honorable mention 2008, and research mandate by the Vrije Universiteit Brussel research council 2012-2022.}

\makebiography{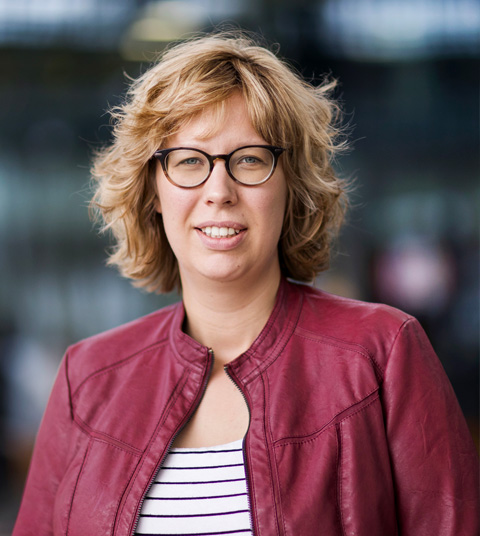}{Sofie Haesaert}{ received the
B.Sc. degree cum laude in mechanical engineering and the M.Sc. degree cum laude in
systems and control from the Delft University of Technology, Delft, The Netherlands, in
2010 and 2012, respectively, and the Ph.D.
degree from Eindhoven University of Technology (TU/e), Eindhoven, The Netherlands, in
2017.
She is currently an Associate Professor with
the Control Systems Group, Department of
Electrical Engineering, TU/e. From 2017 to 2018, she was a Postdoctoral Scholar with Caltech. Her research interests are in the identification, verification, and control of cyber-physical systems for temporal logic
specifications and performance objectives.}
\vfill
\makebiography{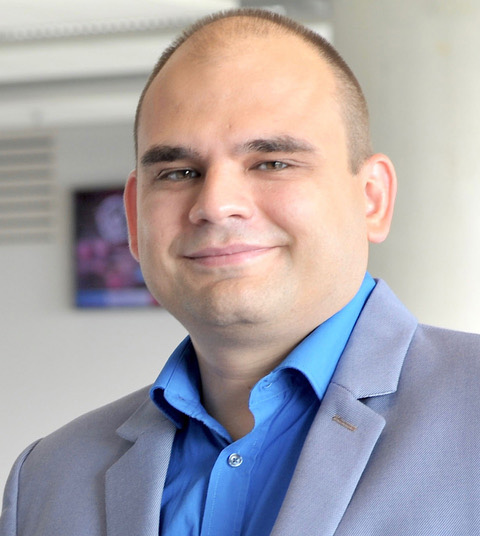}{Roland T\'oth}{received his Ph.D. degree with Cum Laude distinction at the Delft University of Technology (TUDelft) in 2008.  He was a post-doctoral researcher at TUDelft in 2009 and at the Berkeley Center for Control and Identification, University of California in 2010. He held a position at TUDelft in 2011-12, then he joined to the Control Systems (CS) Group at the Eindhoven University of Technology (TU/e). Currently, he is a Full Professor at the CS Group, TU/e and a Senior Researcher at the Systems and Control Laboratory, HUN-REN Institute for Computer Science and Control (SZTAKI) in Budapest, Hungary. He is Senior Editor of the IEEE Transactions on Control Systems Technology and Associate Editor of Automatica. His research interests are in identification and control of linear parameter-varying (LPV) and nonlinear systems, developing data-driven and machine learning methods with performance and stability guarantees for modeling and control, model predictive control and behavioral system theory. On the application side, his research focuses on advancing reliability and performance of precision mechatronics and autonomous robots/vehicles with nonlinear, LPV and learning-based motion control. He has received the TUDelft Young Researcher Fellowship Award in 2010, the VENI award of The Netherlands Organization for Scientific Research in 2011, the Starting Grant of the European Research Council in 2016 and the DCRG Fellowship of Mathworks in 2022.}

\vfill

\end{document}